\documentclass[sigconf]{acmart}

\AtBeginDocument{%
  }
\setcopyright{none} 
\copyrightyear{2025}
\acmYear{2025}
\acmDOI{XXXXXXX.XXXXXXX}

\acmConference[arXiv]{February}{February}{2025}
\acmISBN{$^{*}$Corresponding author: cooperelsworth@google.com}

\usepackage{multirow}
\usepackage{xcolor}
\usepackage{makecell}

\begin{document}

\title{Life-Cycle Emissions of AI Hardware: \\
A Cradle-To-Grave Approach and Generational Trends}
\subtitle{\textit{A Preprint}}
\author{Ian Schneider, Hui Xu, Stephan Benecke, David Patterson, Keguo Huang, \\
Parthasarathy Ranganathan, and Cooper Elsworth$^{*}$ (Google)} 

\begin{abstract}
Specialized hardware accelerators aid the rapid advancement of artificial intelligence (AI), and their efficiency impacts AI's environmental sustainability. This study presents the first publication of a comprehensive AI accelerator life-cycle assessment (LCA) of greenhouse gas emissions, including the first publication of manufacturing emissions of an AI accelerator. 

Our analysis of five Tensor Processing Units (TPUs) encompasses all stages of the hardware lifespan---from raw material extraction, manufacturing, and disposal, to energy consumption during development, deployment, and serving of AI models. Using first-party data, it offers the most comprehensive evaluation to date of AI hardware's environmental impact. We include detailed descriptions of our LCA to act as a tutorial, road map, and inspiration for other computer engineers to perform similar LCAs to help us all understand the environmental impacts of our chips and of AI.

A byproduct of this study is the new metric \textit{compute carbon intensity} (CCI) that is helpful in evaluating AI hardware sustainability and in estimating the carbon footprint of training and inference. This study shows that CCI improves 3x from TPU v4i to TPU v6e. 

Moreover, while this paper's focus is on hardware, software advancements leverage and amplify these gains. 
\end{abstract}

\keywords{Artificial Intelligence, TPU, AI Accelerator, Carbon Accounting, Life-Cycle Analysis, Sustainable Compute}

\maketitle

\section{Introduction}
The energy consumption and environmental impacts of \textit{artificial intelligence} (AI) have generated widespread interest, due to the growing compute requirements of AI models~\cite{Rahman2024}. However, comprehensive assessments of AI's greenhouse gas (GHG) emissions remain limited, hindering the development of standardized quantification methods. This study quantifies the GHG emissions of five Google AI hardware accelerators, adapting the LCA methodology to establish a repeatable process and a consistent metric (\textit{Compute Carbon Intensity} or \textit{CCI}) for wider use.

\begin{figure}[t]
	\centering
	\includegraphics[width=.48\textwidth]{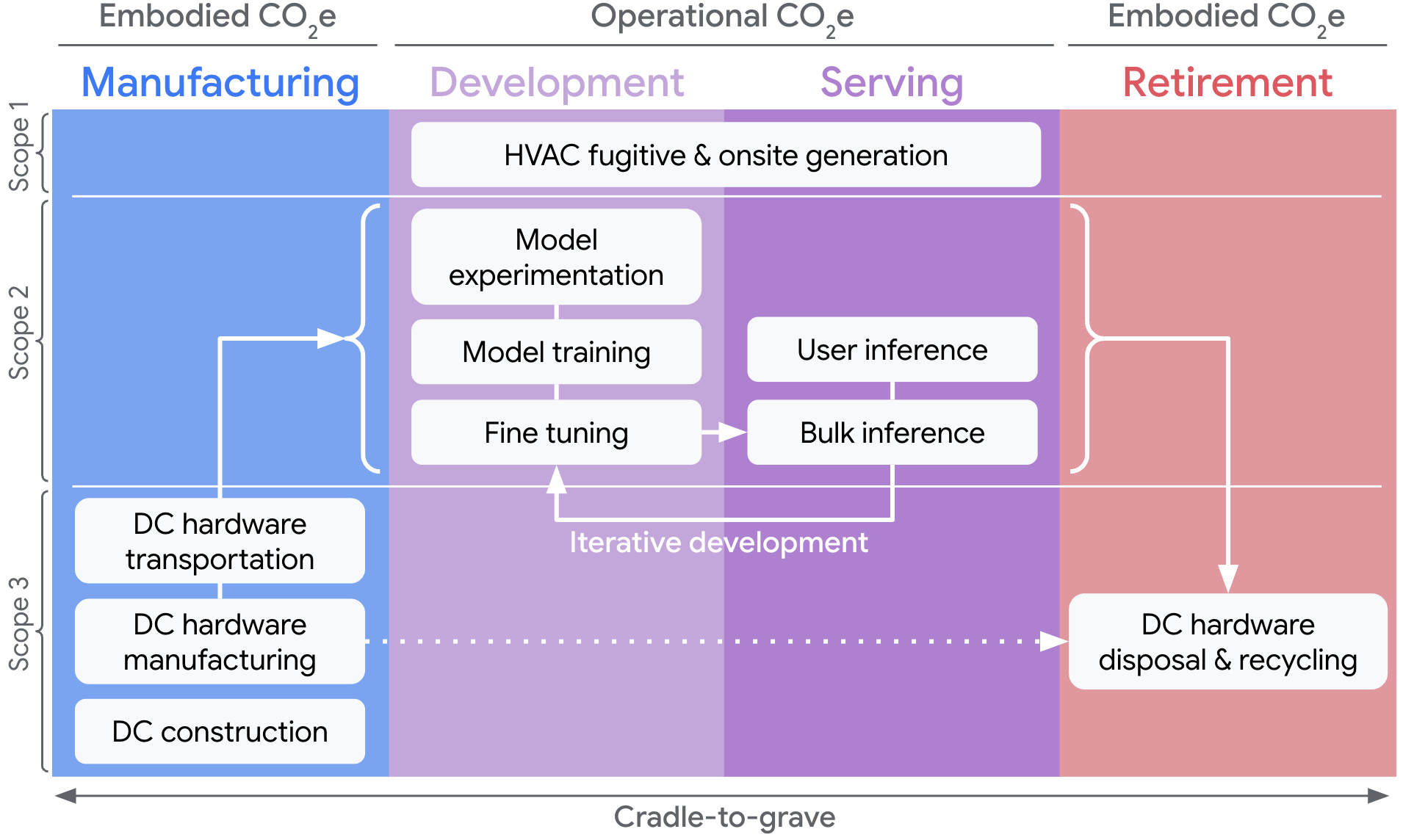}
	\caption{A comprehensive AI cradle-to-grave emissions inventory boundary. It excludes GHG emissions from edge and end-user consumer devices as they are beyond the AI product developer’s control. It omits auxiliary compute and storage emissions sources given the definition of the accelerator-host hardware functional unit.}
	\label{fig:lifecycle-diagram}
\end{figure}

A comprehensive analysis of the GHG emissions associated with AI includes the entire lifespan of the hardware, from the extraction of raw materials to manufacturing, energy use, and eventual disposal. Figure~\ref{fig:lifecycle-diagram} lists the three GHG Protocol ~\citep{GreenhouseGasProtocol2004} emissions sources  vertically---\textit{scopes 1} (direct emissions from sources owned), \textit{2} (indirect emissions from purchased electricity), and \textit{3}  (all other emissions associated with a company)---while also splitting these emissions into three stages across the AI model life-cycle horizontally:
\begin{enumerate}
    \item [(1.a)] \textbf{Data Center Construction.} Emissions associated with data center construction. Data center lifespans are much longer than machine lifespans, so a single data center may serve multiple generations of machines.
    \item [(1.b)] \textbf{Manufacturing and Transportation.} Emissions from the manufacturing of AI hardware and transportation of hardware to data centers.
    \item [(2.a)] \textbf{Operations - Development.} Operational emissions resulting from experimentation, training, and fine tuning of AI models.
    \item [(2.b)] \textbf{Operations - Serving.} Operational emissions associated with bulk and user inference of AI models.
    \item [(3)] \textbf{Retirement.} End-of-life emissions from reverse logistics of retired data center hardware.
\end{enumerate}

\textit{Embodied emissions} refers to stages 1 and 3:  GHG emissions associated with upstream (extraction, parts production, server manufacturing, and transportation) and downstream (end-of-life, and recycling) stages of the data center hardware life-cycle. \textit{Operational emissions} refers to stages 2.a and 2.b: GHG emissions associated with usage of the data center hardware.

\textit{Tensor Processing Units} (TPUs) are Google's AI-focused \textit{applica- tion-specific integrated circuits} (ASICs). These ASICs offer substantial improvements in training and inference speeds for large-scale machine learning models. Table ~\ref{tab:specs} lists characteristics of the five TPUs in this study.

\begin{table*}[ht] 
    \caption{TPU Features. The first segment lists hardware specifications, the second segment provides fleet power measurements, and the LCA result (CO$_{2}$e emissions over hardware lifetimes) is presented in  the third segment.  Section ~\ref{section:definition} defines market-based (MB) and location-based (LB) emissions. The fourth segment shows \textit{Compute Carbon Intensity} (CCI), embodied CCI, and operational CCI (for MB and LB). Lower is better. Unless noted, we normalize data per TPU chip, but include a machine's entire power and emissions, including non-TPU items like a CPU.}  
    \centering
\begin{tabular}{| c | p{5.2cm} | c c c | c c|}
\hline
 & {TPU version}   & TPU v4i~\citep{Jouppi2021}  & TPU v5e\citep{semiTPUv5e} & TPU v6e~\citep{Vahdat2024} & TPU v4~\citep{Jouppi2023}   & TPU v5p~\citep{Vahdat2023}  \\
  & TPU type (number/CPU host server)   & \multicolumn{3}{ c |}{Versatile TPUs (8/server)}  &  \multicolumn{2}{ c |}{Powerful TPUs (4/server)}  \\
 \hline
\multirow{6}{*}{1. TPU specs} & Production deployment & 2020  & 2023 & 2024 & 2020  & 2023  \\
 & Peak  TeraFLOP/second (BF16)                 & 138                                    & 197                          & 918                                  & 275                                  & 459                                    \\
 & Die size (mm$^{2}$), Technology &<400 , 7nm                &  N.A.\footnotemark                & $\textrm{N.A.}^1$                                 & <600, 7nm             &  $\textrm{N.A.}^1$                                   \\
 & HBM type, capacity (GiB)                           & HBM2, 8                                   & HBM2E, 16                        & $\textrm{N.A.}^1$, 32                                 & HBM2, 32                                 & HBM2E, 96                                   \\
 & HBM bandwidth                   & 614 GB/s                        & 819 GB/s             & 1640 GB/s                    & 1200 GB/s                    &  2765 GB/s  \\
 & CPU Host DDR DRAM capacity (GiB)                 & $\textrm{N.A.}^1$                                & 512                       & 1536                              & $\textrm{N.A.}^1$                              & $\textrm{N.A.}^1$                                \\
\hline
\multirow{3}{*}{\makecell{2. Measured \\ Power and \\ Effectiveness}} & Mean power (W/machine) & 1,184 & 1,171 & 2,173 & 1,167 & 2,176 \\
 & Mean power (W/TPU excluding host) & 75 & 66 & 153 & 164 & 331    \\
 &  \textcolor{black}{Kilowatt-hours per $10^{18}$ FLOPs} &  \textcolor{black}{2.53} &  \textcolor{black}{2.16} &  \textcolor{black}{0.86} &  \textcolor{black}{1.93} &  \textcolor{black}{1.65} \\
\hline
\multirow{5}{*}{\makecell{3. Life-cycle \\ Emissions \\ (kgCO$_{2}$e \\ \textcolor{black}{over 6 years)}  } } & Embodied CO$_{2}$e (kgCO$_{2}$e)  & \textcolor{black}{386} & \textcolor{black}{402} & \textcolor{black}{692}  & \textcolor{black}{693} & \textcolor{black}{1,101}  \\
&  \multicolumn{1}{r |}{(Data Center Construction)}       & \multicolumn{1}{r}{\textcolor{black}{(59)}} & \multicolumn{1}{r}{\textcolor{black}{(59)}} & \multicolumn{1}{r |}{\textcolor{black}{(109)}}   & \multicolumn{1}{r}{\textcolor{black}{(117)}}  & \multicolumn{1}{r |}{\textcolor{black}{(218)}} \\
 & \multicolumn{1}{r |}{(CPU Manufacturing+Transportation)} & \multicolumn{1}{r}{(119)}  & \multicolumn{1}{r}{(106)} & \multicolumn{1}{r |}{(260)} & \multicolumn{1}{r}{(210)} & \multicolumn{1}{r |}{(299)} \\
 & \multicolumn{1}{r |}{(TPU Manufacturing+Transportation)} & \multicolumn{1}{r}{(208)}  & \multicolumn{1}{r}{(238)} & \multicolumn{1}{r |}{(323)} & \multicolumn{1}{r}{(366)} & \multicolumn{1}{r |}{(585)} \\
 & Operational CO$_{2}$ market-/location-based & \textcolor{black}{1166/3137} & \textcolor{black}{1154/3104} & \textcolor{black}{2141/5759}  & \textcolor{black}{2301/6187} & \textcolor{black}{4288/11532}   \\
\hline
\multirow{2}{*}{\makecell{4. CCI (gCO$_{2}$e \\ /$10^{18}$ FLOPs)}} & Embodied CCI (gCO$_{2}$e  \textcolor{black}{per $10^{18}$ FLOPs})    &   \textcolor{black}{114}  & \textcolor{black}{103}  & \textcolor{black}{38} &       \textcolor{black}{79}    &  \textcolor{black}{58}   \\
 & Operational CCI market-/location-based   & \textcolor{black}{346/929}   & \textcolor{black}{295/793} & \textcolor{black}{118/316}  & \textcolor{black}{263/709}   & \textcolor{black}{225/605}   \\
\hline                               
\end{tabular}
\label{tab:specs}
\end{table*}

The distinction between TPUs available through Google Cloud Platform follows their suitability for different tasks:
\begin{itemize}
    \item \textit{Versatile TPUs} are cost-efficient options for broad accessibility, aiding both training and inference of large AI models with optimizations for inference of modern AI models~\citep{Vahdat2023}. They include TPU v4i, TPU v5e, and TPU v6e (Trillium).
    \item \textit{Powerful TPUs} are high-performance, focusing on fast LLM training, boasting high FLOP performance, and high-bandwidth memory. They include TPU v4 and TPU v5p.
\end{itemize}
TPUs may be more energy efficient than typical AI accelerators since they rely on 1 or 2 processors with very large systolic arrays versus numerous processors with small multiplier arrays ~\citep{jouppi2017datacenter}.

To reduce confusion about the plural of the abbreviation FLOP (floating point operation), this paper \textit{never} uses FLOPS to represent FLOPs per second. Performance rates have ``/second'' appended (e.g., TeraFLOP/second) while a quantity of arithmetic operations have a number or prefix beforehand (e.g., $10^{18}$ FLOPs or 1 ExaFLOP mean one quintillion floating point operations).
\footnotetext{Not available (N.A.) are values not disclosed. We use these values internally to compute manufacturing emissions here and in Figure~\ref{fig:manufactuing-emissions}. Fortunately, publishing manufacturing emissions does not inherently expose proprietary design information. We hope this approach motivates more designers to complete and disclose similar LCA analyses.}

\textbf{Contributions:} First, this paper estimates comprehensive cradle-to-grave emissions of AI hardware, presenting emissions impacts using first-party data. We specifically target AI accelerators and the attached host computer. This \textit{life-cycle analysis (LCA)} can directly inform efforts to quantify and reduce the carbon emissions of services offered in data centers on specialized AI hardware. 

Second, the paper proposes a solution to the open problem of how to quantitatively and fairly compare AI machines with differing performance and carbon footprints by defining a new normalized CO$_{2}$e / performance metric for AI hardware: \textit{compute carbon intensity (CCI)}. For CCI, lower is better: it means lower carbon emissions for the same amount of computation. We also present operational and embodied versions of CCI that can match narrower evaluations. 

Third, as part of the LCA, this paper provides an accurate and replicable measurement of operational emissions of AI hardware, using direct measurement across Google's entire modern fleet of TPUs at a snapshot in time. This approach improves on estimating operational emissions based on machine thermal design power (TDP), because actual operating energy and emissions vary substantially depending on utilization and software, e.g., deployed TPU v4s can vary by 60\%~\citep{Jouppi2023}. Our approach also naturally captures heterogeneity in AI hardware utilization and power consumption, rather than using a single model. 

To date, studies of AI products have largely focused on only one or two life-cycle stages. As part of our LCA, this paper is the first  to estimate manufacturing emissions of AI accelerators. Prior work used proxies like 150 kg of emissions based on consumer hardware (see Section~\ref{section:related}). The Appendices document our LCA process including the tools used, which we hope will act as a tutorial, road map, and inspiration for others to publish many more LCAs.

We also directly compare the operational emissions and embodied emissions of successive generations of AI hardware. This longer term view gives new insight into the likely improvement in energy and carbon efficiency of AI hardware over time, which is critical for understanding and forecasting AI's carbon impacts. 

Finally, building on the six contributions above, we directly measure a massive collection of AI hardware accelerators in production across a wide number of hyperscale data centers to calculate actual achieved CCI. This rigorous methodology establishes that CCI improves 3x from TPU v4i to TPU v6e. This rapid gain---in just 4 years---demonstrates the effectiveness of Google's hardware design process and manufacturing efficiency improvements.

\section{Definition of concepts and assumptions}
\label{section:definition}
The functional unit for this study is one AI computer deployed in the data center, which includes one or more accelerator trays (containing TPUs) connected to one host tray (i.e., a computing server). This configuration is a standard practice across all Google TPU generations. Due to the definition of the functional unit, peripheral components beyond the tray (e.g., rack, shelf, network equipment) and auxiliary computing and storage resources are excluded from the calculation of embodied and operational emissions. We include data center cooling overheads in operational emissions.
 
We use a lifespan of six years for AI machines. While machine lifespans vary, this length is a reasonable assumption in line with those found previously~\citep{Vahdat2024}. Appendix \ref{appendix:steps} provides more details on the specific aspects of the AI computer's lifecycle that are included in the study, using the GHG Protocol~\citep{GreenhouseGasProtocol2004}---the internationally accepted GHG accounting and reporting standard---to define the categories of the emission inventory boundary. 

For all platforms covered in this study, GHG emissions are in grams or kilograms of carbon dioxide equivalent emissions (gCO$_{2}$e or kgCO$_{2}$e). We use \textcolor{black}{emission factors} from the IPCC Fifth Assessment Report (AR5)~\citep{Myhre2013} to estimate GHG emissions associated with data center operations. While newer factors may be available, the AR5 provides a widely recognized and standardized methodology, ensuring comparability with other studies and facilitating a consistent assessment of AI hardware's climate impact.

To calculate electricity-related emissions  (``Scope 2''), we multiply operational electricity consumption by the annual average \textcolor{black}{electricity \textit{emission factor}}, which can be assessed in a number of ways. Two commonly accepted methods come from the GHG Protocol:
\begin{itemize}
    \item \textit{Location-based} (LB) GHG emissions refer to the GHG emissions emitted within a specific geographic boundary, such as a country or grid region. LB emissions are calculated using the annual average \textcolor{black}{electricity emissions factor} for the defined grid. They exclude the impact of \textit{carbon-free energy} (CFE) procurement, so it's the most conservative standard we consider.
    \item \textit{Market-based} (MB) GHG emissions account for the GHG emissions emitted by generating sources from which a company purchases electricity and associated environmental attributes. It thus credits companies for CFE purchases, allowing them to reduce their associated MB emissions. \textcolor{black}{Today’s Scope 2 Standard allows companies to reduce their MB emissions by purchasing CFE to match their demand on an annual basis within broad geographic boundaries that may not reflect deliverability of that electricity.}\footnote{\textcolor{black}{For example, the United States and Canada are treated as one market boundary, even though both countries comprise many different grids and balancing authorities. Similarly, Europe is considered as one market boundary, even though it contains many different electricity pricing zones and each country has its own transmission system operator (TSO) that plans and operates the national electricity system.}} 
\end{itemize}
In addition to the primary GHG Protocol-aligned reporting, in Section \ref{result: lca} we discuss electricity emissions factors, emissions, and CCI under a 24/7 hourly CFE matching standard~\citep{google24x7carbonfree} that researchers have found to be more effective than MB accounting~\citep{xu2024system,riepin2024means,IEA2022}.
The relationship between MB and LB accounting is
\begin{equation*}
   \begin{aligned}
        \textcolor{black}{\textrm{MB emission factor}} &=  \textcolor{black}{\textrm{LB emission factor}} \\ 
        & - \textcolor{black}{\textrm{Impact of procured CFE }}
    \end{aligned}
\end{equation*}
\textcolor{black}{For 2023, Google's LB emission factor was 366 gCO$_{2}$e/kWh and the reduction associated with its procured CFE was equivalent to 231 gCO$_{2}$e/kWh, so the net MB emission factor was 135 gCO$_{2}$e/kWh.} \textcolor{black}{We multiply each TPU's estimate of annual machine energy consumption by these Google-wide annual average \textcolor{black}{electricity emission factors} to calculate operational MB and LB emissions.} 

Table \ref{tab:specs} reports \textcolor{black}{emission factors} under \textcolor{black}{both LB and MB} accounting methodologies to illustrate how accounting choices impact emissions reduction interventions.  LB is  essential because it allows for the most direct comparison of AI machine energy consumption; other metrics may obfuscate hardware-specific differences as CFE procurement can vary by time and operator. MB is valuable because it credits CFE procurement based on today’s Scope 2 Standard, which mitigates operational electricity emissions. Disclosing more information seems wise given the uncertainty about a long-term industry standard (see Section \ref{result: lca}).

While we aspired to be exhaustive, Appendix Section \ref{appendix:exclusions} clearly explains exclusions from our LCA. Appendix Sections \ref{appendix:Emissions} to \ref{appendix DC} list the standard accounting practices we follow for calculating manufacturing emissions, operational emissions, end-of-life emissions, and data center construction emissions.   We followed ISO 14040/14044 to assess comprehensive cradle-to-gate emissions. To model sophisticated chips like TPUs, we  combine  advanced life-cycle inventories for front-end processes from IMEC's virtual fab with proprietary chip manufacturing parameters---e.g., tech node, die sizes, yield rates, fluorinated GHG abatement ratios---for back-end chip and panel-level assembly.
Our end product is the \emph{first publication of manufacturing emissions of AI accelerators}.

For those interested in learning more about computer hardware LCAs, the appendices may act as a tutorial. Readers disinterested in LCA details can skip the appendices and simply proceed to the next section on performance metrics.

\section{CCI: A novel CO$_{2}$e/performance metric}
\label{section:methods}
~\citet{vahdat2024new} opine that we need new metrics to guide us ``to find a path to grow AI and cloud computing efficiently and responsibly.'' To accurately compare the emissions across generations and relative to other architectures, we need to define a suitable measure of \textcolor{black}{carbon intensity}. ~\citet{vahdat2024new} suggest a performance measure of CO$_{2}$e relative to ``Workload Goodput'': CO$_{2}$e/Goodput.

We propose a new metric related to CO$_{2}$e/Goodput based on first-party measurement data of workloads run in data centers that we call \textit{compute carbon intensity} (CCI). It is the CO$_{2}$e per utilized \textit{floating-point operation} (FLOP), or CO$_{2}$e/FLOP. 
\textcolor{black}{\textit{The denominator of CCI is a fixed amount of computation, not a rate: its unit is number of FLOPs, not FLOPs/second.}}

This assessment aspires towards the CO$_{2}$e/Goodput ideal by accounting for the varying throughput and energy use of accelerators based on evaluations in the fleet and on their embodied footprint, achieving something close to the ideal CO$_{2}$e/Goodput metric that can be based on real measurements.

~\citet{vahdat2024new} explain three reasons why goodput is superior to existing metrics for performance normalization: 
\begin{enumerate}
    \item Running the actual workload is more accurate than running a benchmark. Our metric uses power and performance data from actual workloads deployed across the AI fleet. 
    \item Goodput adjusts workload performance by subtracting effort wasted on underutilization. Our proposed metric does that naturally. When machines are underutilized, their power is still significant; they use $\sim$60\% of their average power when idle~\citep{Fan2007}. Underutilized machines have worse CO$_{2}$e/FLOP.
    \item \citet{vahdat2024new} suggest that a useful goodput adjustment is to remove wasted work due to unreliability. Our proposed metric doesn't feature that final adjustment. We focus on the hardware functional unit, and a reliability correction would require additional details on workload reliability that could obfuscate the differences in CO$_{2}$e intensity across different AI hardware generations. Incorporating measures of wasted work due to unreliability could be useful for CO$_{2}$e / performance metrics in the future. 
\end{enumerate}

One benefit of CCI is that if we know the approximate number of floating point operations performed by an AI hardware accelerator for a training run or for an inference, we can use CCI to get a ballpark estimate of its carbon emissions. For example, training GPT-3 takes roughly $3.14 \times 10^{23}$ floating-point operations~\citep{Patterson2021}.  As shown in Table \ref{tab:specs}, CCI for TPU v4 is about 342 gCO$_{2}$e/ExaFLOP ($10^{18}$ floating-point operations) and about 283 gCO$_{2}$e/ExaFLOP for its successor TPU v5p. Training GPT-3 on TPU v4 would result in emissions of approximately $342 \times 3.14 \times 10^{23}/10^{18} = 10.7 \times 10^{7}$ grams or 107 tonnes of CO$_{2}$e.
For TPU v5p, the ballpark estimate for training GPT-3 is 89 tonnes of CO$_{2}$e, a savings of nearly 20\%.\footnote{We pick ExaFLOP ($10^{18}$ FLOPs) so that the numerator can be in grams of CO$_{2}$e. One \textit{tonne} is 1000 kg, or a \textit{metric ton} in the US. For GPT-3 on TPU v4, 107 tonnes of CO$_{2}$e = 25 tonnes of embodied CO$_{2}$e + 82 tonnes of market-based operational CO$_{2}$e.}

\textcolor{black}{Segment 4 of Table ~\ref{tab:specs} shows that we can divide CCI into embodied and operational components. What is the relationship between Operational CCI and the traditional performance/Watt metric? If Performance is average utilized FLOPs/second and Watt (Joules/second) is the measured, whole-system, average power (not just the ASIC), then Performance/Watt is also FLOPs/Joule and
\begin{equation*}
     \textrm{Operational CCI} = \frac{\textrm{Electricity Emissions Factor}} {\textrm{Performance}/\textrm{Watt}}
   \end{equation*}
One advantage of Operational CCI is the ease of including embodied carbon in total CCI, which is difficult for performance/Watt. A second advantage is the ability to account for efforts to reduce carbon emissions by siting data centers in locations with low-carbon grids or by buying more CFE.}

\textcolor{black}{To measure Operational CCI, CO$_{2}$e/FLOP, the first step is collecting \textit{utilized FLOPs}. This denominator of CCI is calculated from an internal measurement that provides the total number of floating-point operations (FLOPs), for each chip, for each five-minute interval. We can sum FLOPs over chips and five-minute intervals to get the total number of FLOPs for any machine or collection of machines for the time period of interest.}

\textcolor{black}{To measure progress in CCI for similar TPUs--versatile TPUs or powerful TPUs---one concern is how to  account for potential bias due to varying TPU utilization.  Higher utilization leads to better CCI, as idle machines consume power but don't compute.  We thus measure \textit{Duty Cycle} of machines in the fleet. It is a chip-level utilization metric representing the fraction of time a chip is in use. It is a value between 0.0 to 1.0, e.g., if duty cycle = 0.5, then the chip is in use for half of the interval. The duty cycle for a machine is the average duty cycle across all chips in the machine.} 

\textcolor{black}{How does one adjust CCI for similar TPUs based on their average duty cycles? We employ \emph{propensity score weighting} ~\citep{Rosenbaum1983} to estimate CCI based on a sampled duty cycle distribution approximating the fleetwide average for comparable TPUs. This fairness statistical technique has a modest but important effect for two of the five CCI values;  for example, it reduces the measured CCI improvement of v6e over v4i from 74\% (in the raw data) to 66\% and v5p over v4 from 24\% to 16\%; v5e remains at 14\% over v4i for both versions. Propensity score gives a slightly more accurate picture of how much the hardware itself has improved, independent of how it is being used in the fleet} 
(see Appendices \ref{appendix: benchmark_power} and \ref{appendix:Propensity} for details).

Some machines have missing data. They may have power readings but lack utilization or performance metrics, for example, if a machine was just deployed. To avoid skewing the analysis, these machines are excluded.

Like most AI chips, TPUs offer several  floating-point formats of different widths, with the narrow ones being faster \textit{and} using less (operational) energy. To reduce arguments, we define CCI as being based on the \textit{measured} number of FLOPs using runtime machine counters and the \textit{measured} energy consumed for whatever the resulting mix of data widths is for \textit{workloads on deployed machines}. \textcolor{black}{A downside of treating all computations equally is that identifying the impact of narrow data types is hard. Future work can evaluate the quality and CO$_{2}$e trade-offs of varying numerical formats. }

\textcolor{black}{Our use of effective FLOPs aligns well with the AI community’s use of it to measure algorithmic efficiency.\footnote{\citet{Schwartz2020} advocate using FLOPs to measure the efficiency of AI models.} Like all performance metrics, it has limitations; e.g., it implicitly anchors to the current state of algorithms in current workloads. For example, say a new algorithm requires 10\% fewer FLOPs to get the same result. Multiplying FLOPs by CCI suggests this algorithm reduces carbon emissions by 10\%. Since CCI  is based on the average power and computation of workloads in the fleet, if one reduced the computation requirements of \textit{all} workloads by 10\%---decreasing both energy and measured effective FLOPs---it would ideally render CCI unchanged. However, in practice, we might see measurable results depending on the relative impact on energy and effective FLOPs.}  

\section{Results and discussion}

\subsection{Result 1: Generational trends of TPU \textcolor{black}{carbon intensity}}\label{result:2}
This result shows how the life-cycle \textcolor{black}{carbon intensity} of Google TPUs, normalized by performance, has improved with each successive machine generation. It is essential to normalize by performance because newer machines are substantially faster than older ones. In fact, the  peak floating-point operations (BF16) per second  of a single TPU v6e matches that of 5 to 7 TPU v4is or TPU v5es. 

As mentioned above, to reduce bias from different utilization of successive TPU generations,  we use \textit{propensity score weighting}. This balances the distribution of duty cycle utilization for similar TPUs. See Appendix \ref{appendix:Propensity} for more information. From the balanced data, we calculate, for each machine and five-minute interval: (a) carbon emissions (gCO$_{2}$e) and (b) total computations, measured in floating point operations.

Calculating the ratio of (a) and (b) yields compute carbon intensity (CCI): gCO$_{2}$e/ExaFLOP. This metric allows for a robust comparison between accelerators and AI workloads, showcasing efficiency improvements over time.

\begin{figure}[t]
	\centering
	\includegraphics[width=.48\textwidth]{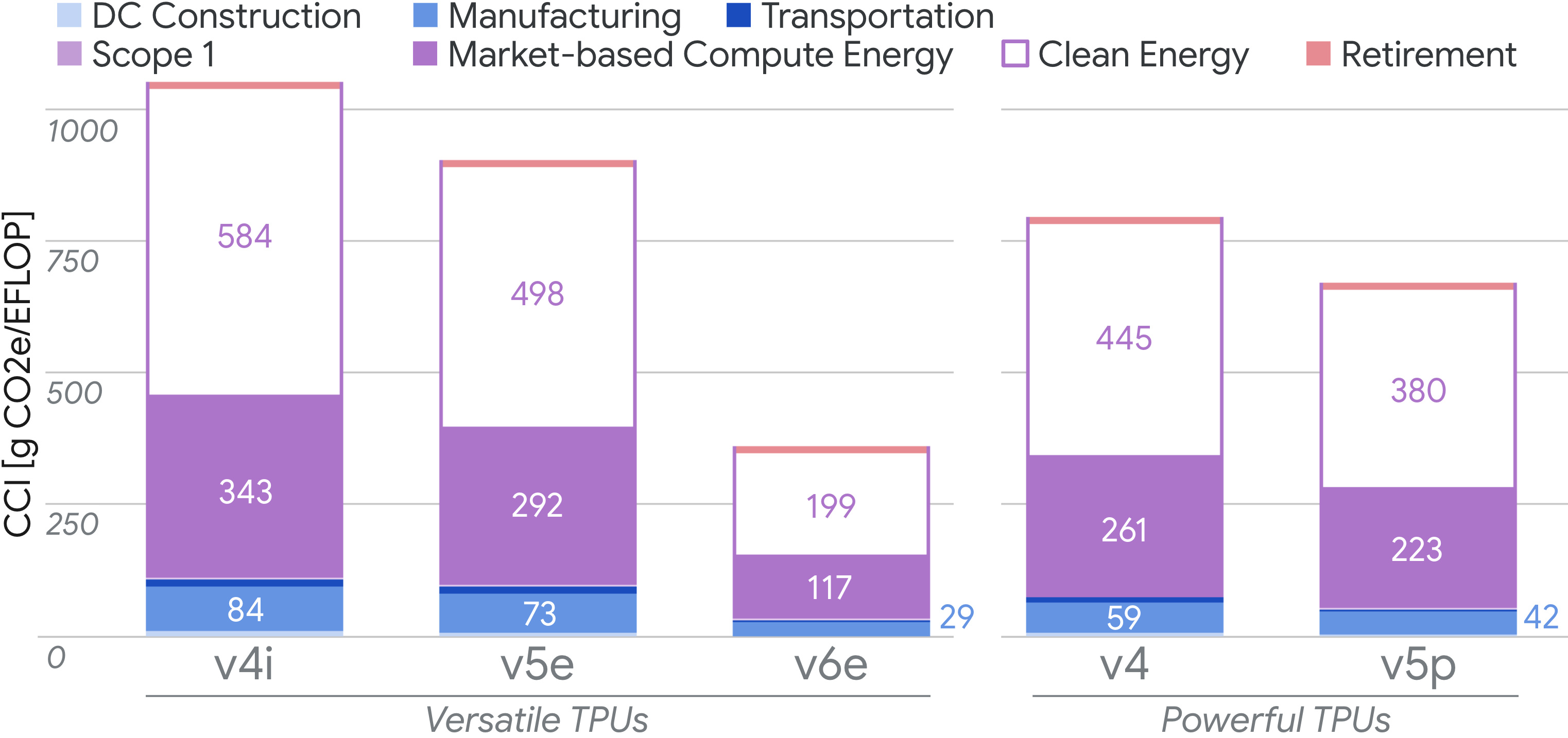}
	\caption{Lifetime AI hardware compute carbon intensity (CCI), with a breakdown by life-cycle stage.}
	\label{fig:lifetime-emissions-intensity}
\end{figure}

Figure~\ref{fig:lifetime-emissions-intensity} summarizes the results using the CCI metric. Within a TPU class, successive generations show dramatic improvements in CCI. Trillium (TPU v6e), launched to customers in December 2024, offers a 3x improvement on CCI versus v4i. The quadrupling of its systolic array size explains in part v6e's large gain.

\begin{figure}[t]
	\centering
	\includegraphics[width=.48\textwidth]{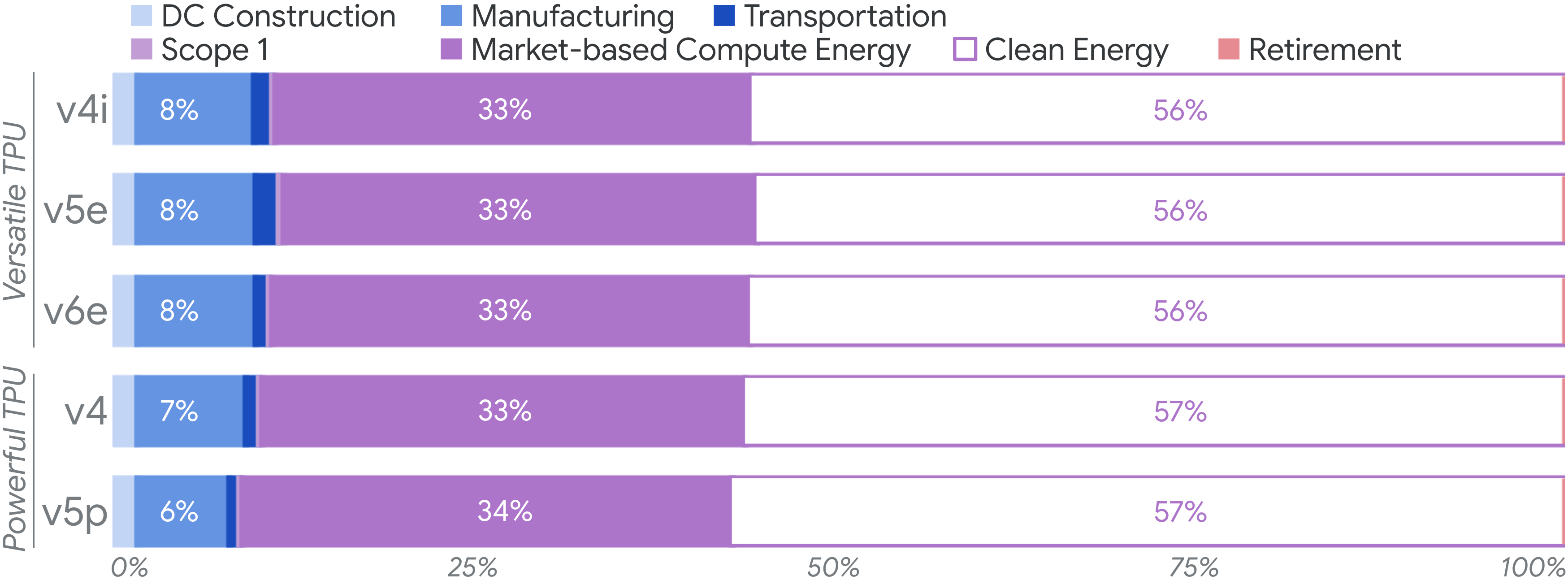}
	\caption{Accelerator and host emissions by life-cycle stage. 
	}
	\label{fig:rule-of-thumb}
\end{figure}

\subsection{Result 2: Emissions by life-cycle stage} \label{result: lca}
The life-cycle GHG emissions methodology presented in the appendices allows us to completely characterize the life-cycle emissions of deployed AI hardware. The third segment of Table~\ref{tab:specs} above summarizes lifetime emissions for Google’s several recent generations of TPUs, and Figure~\ref{fig:rule-of-thumb} shows the percentage break down of emissions by contributions from each life-cycle stage. 

These results illustrate that over the six year accelerator lifetime:
\begin{itemize}
    \item Operational electricity emissions dominate embodied hardware emissions. Operational emissions are $\sim$70\% of emissions when including Google’s CFE procurement (market-based electricity) and  $\sim$90\% of emissions when not including CFE procurement (location-based).
    \item Google’s CFE purchases substantially mitigate operational emissions \textcolor{black}{under the GHG-protocol aligned MB operational emissions metric}, reducing them by more than half. 
\end{itemize}

\textcolor{black}{
\textbf{Consider more accurate standards for operational emissions accounting.} 
We can more accurately account for operational emissions and more effectively mitigate them through more selective CFE purchases.  
The current market-based (MB) approach under the GHG Protocol allows for emissions mitigation strategies that differ widely in effectiveness.\footnote{\textcolor{black}{
The GHG protocol allows companies to match their MWh of consumption with purchases of MWh of clean energy, substantiated through the use of \textit{Energy Attribute Certificates} (EACs). Some companies purchase EACs and electricity together (“bundled EACs”). Some purchase EACs without purchasing the underlying CFE that it represents (“unbundled EACs”). Current practice allows EACs purchased anywhere within broad geographic boundaries (e.g. all of North America or all of Europe) and at any time of year to be used to reduce MB emissions, a practice that has come under increasing criticism~\citep{lott2023rethinking}. Google typically signs long-term contracts for clean power and EACs, and has strong additionality principles for our procurement~\citep{google2024corporate}. Google has also been working to advance time-based Energy Attribute Certificates (T-EACs) which offer a more credible instrument to reduce MB emissions. Some companies' emissions claims rely heavily on the use of unbundled EACs that are disconnected from underlying electricity consumption~\citep{peters2024}.
}}
Most companies focus on procuring CFE to match their annual electricity demand, often from projects located far from where they consume power. 
A significant and growing body of academic research employing robust energy system modeling approaches has shown that corporate matching of electricity use on a locational and hourly basis, even at levels below 100\%, can drive greater grid decarbonization and accelerated technology innovation compared to matching electricity demand at an annual level~\citep{xu2024system,riepin2024means,IEA2022}.
}

\textcolor{black}{
In 2020, Google set an ambitious goal to run on 24/7 CFE on every grid where we operate by 2030~\citep{google24x7carbonfree}. This approach involves matching electricity consumption with CFE sources on an hourly basis, ensuring that operations are met with clean energy at all times. To better align with this goal, we can calculate electricity emissions factors using a 24/7 methodology that employs stricter geographical and temporal matching requirements for CFE; see Appendix \ref{appendix:24/7} for more details. This methodology provides a more accurate and conservative picture of emissions reductions from CFE procurement compared to the predominant MB approach (i.e., the electricity emission factor will be higher).
}

\textcolor{black}{
Under the hourly 24/7 standard, Google's global 2023 electricity emissions factor would be 212 gCO$_{2}$e/kWh instead of 135 gCO$_{2}$e/kWh under MB accounting. Applying the 24/7 methodology to our TPU v6e analysis reveals that lifetime electricity emissions would be 3,305 kg, and the operational CCI would be 182 gCO$_{2}$e/ExaFLOP. Under 24/7 accounting, operational emissions account for 82\% of total v6e emissions, compared to 75\% under MB accounting. The 24/7 accounting standard more accurately accounts for CFE purchases, which emphasizes the importance of aligning clean energy purchases with actual electricity demand to reduce the carbon intensity of AI hardware. 
}

\textcolor{black}{
\textbf{What might the future hold?} To explore the potential impact of increased CFE penetration, we consider a hypothetical scenario where each of Google’s data centers around the world are matched with 90\% CFE on a local and hourly basis. This scenario is inspired by Google's progress in several regions where CFE already exceeds 90\%. 
For those 90\%+ regions, the load-weighted average 2023 electricity emissions factor under the hourly 24/7 standard is 31 gCO$_{2}$e/kWh. We can use this to approximate future emissions with 90\% CFE under the stricter hourly 24/7 standard. In a future scenario with 90\% CFE globally, under the hourly 24/7 standard, the lifetime electricity emissions for a TPU v6e would be around 509 kg, and the operational CCI would be around 27 gCO$_{2}$e/ExaFLOP. Under these conditions, operational emissions would account for around 40\% of total v6e emissions. Attaining 90\% CFE improves the total CCI (operational plus embodied) of v6e by 3.3x. This result highlights the benefits of increased CFE procurement.
}

\textcolor{black}{
In a future world where both use \textit{and manufacturing} is powered by energy with 90\% CFE, emissions would be significantly reduced.  (Current CFE scores are much worse in the Asia Pacific region where most manufacturing emissions occur, so this scenario is even more hypothetical.) For example, assume that 50\% of manufacturing emissions are due to electricity consumption (\citet{TSCM_SR} reports that about 50\% of their emissions today are from Scope 2, but it could change if reporting for upstream categories expands). 
If we assume that TPU manufacturing sites are also at 90\% CFE, this reduces manufacturing emissions by 47\% (the reduction is amplified because of a worse 2023 CFE score in Asia-Pacific) and improves embodied CCI by an additional factor of 2x. Attaining 90\% CFE globally, reducing operational and manufacturing emissions, improves the total CCI (operational plus embodied) of v6e by 4.6x (under the hourly 24/7 accounting standard). In this potential futuristic world operational and embodied emissions would each be around 50\% of total lifecycle emissions.  
}

\textcolor{black}{
Combining hardware improvements with 90\% CFE scenarios: a TPU v6e with 90\% CFE operational emissions has a 10x improvement in CCI (under the hourly 24/7 accounting standard) relative to a TPU v4i, and a v6e with 90\% CFE operational emissions \textit{and} 90\% CFE for electricity-related manufacturing emissions has a 14x improvement in CCI relative to a v4i.
}

\textcolor{black}{
\textbf{How should a computer architect weigh operational and embodied emissions?}
From a computer architect's perspective, the accounting methodology and projected rate of CFE deployment and procurement have a big impact on the relative importance of operational versus embodied emissions.  In our view, however, the primary lesson remains fairly consistent across each methodology. Ignoring CFE procurement, a rule-of-thumb is that embodied emissions are $\sim$10\% and operational emissions are $\sim$90\% of an AI system’s lifetime emissions.  To attack the biggest slice of the pie, work to reduce operational emissions by improving energy efficiency (Perf/W) or procuring CFE. Reducing embodied emissions becomes more important as operational emissions decline, but it will be more challenging than operational emissions because it would require coordinated efforts across the supply chain.
}

\textcolor{black}{As operational emissions are reduced (e.g., CFE procurement), the relative impact of embodied emissions becomes larger under either the MB or 24/7 accounting methodologies. Nevertheless, under the more accurate 24/7 accounting, operational emissions remain important at around 50\% even if CFE were to grow to 90\%. 
}

\subsection{Result 3: Manufacturing emissions trends by TPU generation}\label{result:3}
Result \ref{result:2} illustrates that CCI is decreasing with each successive generation of AI hardware. That trend is replicated when we specifically consider machine manufacturing emissions (blue parts of Figure~\ref{fig:lifetime-emissions-intensity}). The embodied CCI decreases by 66\% from v5e to v6e and by 29\% from v4 to v5p. The deep reduction from v5e to v6e is because peak performance improved by 4.7x whereas manufacturing emissions increased by only 1.8x. While new TPUs are more carbon intensive to manufacture and operate, the new machines provide much more useful workload throughput than older AI machines.

\begin{figure*}[t]
	\centering
	\includegraphics[width=.85\textwidth]{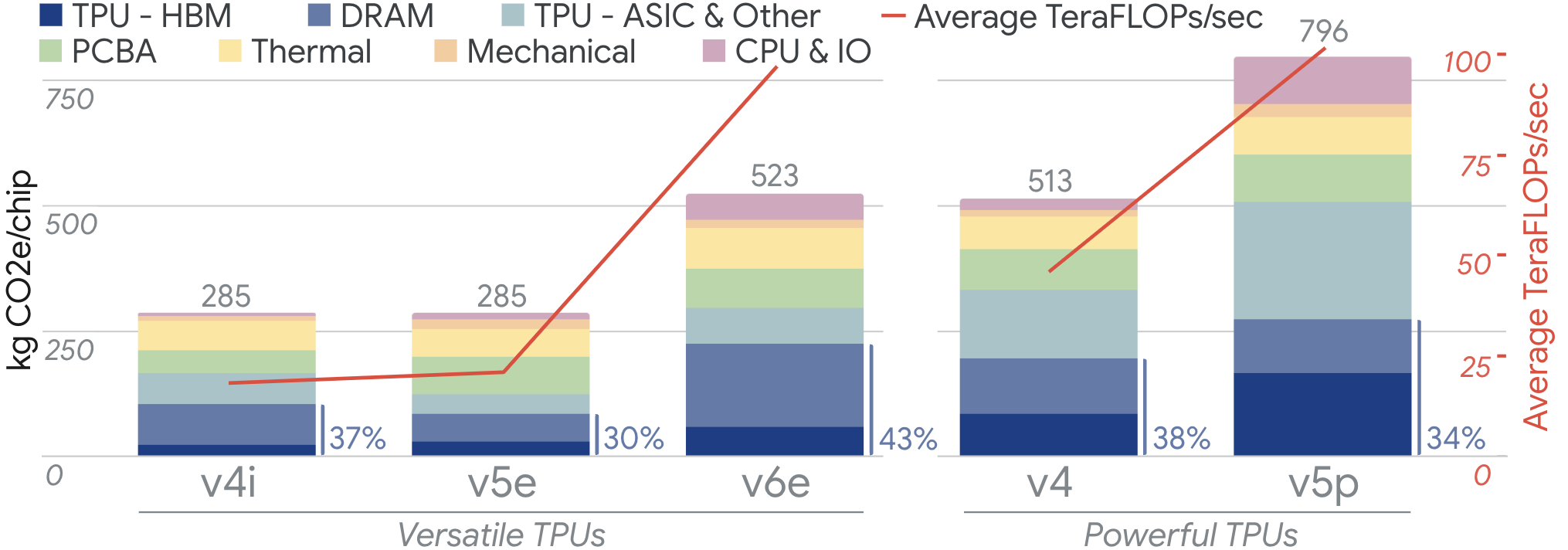}
	\caption{Absolute manufacturing emissions per chip, by generation and by component type, and average TeraFLOP/second. It shows the total manufacturing emissions per machine, normalized by the number of chips per machine. (Table~\ref{tab:specs} embodied includes DC construction and transportation emissions.) The memory piece of this CO$_{2}$e (TPU HBM + CPU DRAM) averages $\sim$38\%.  
	}
	\label{fig:manufactuing-emissions}
\end{figure*}

Figure~\ref{fig:manufactuing-emissions} on the next page shows the increase in absolute manufacturing emissions of newer generations and highlights the primary component drivers of the increase in absolute emissions. The adoption of advanced manufacturing technologies and more resources added to the machine both increase absolute emissions.

Taking the three versatile TPU platforms (i.e., v4i, v5e, v6e), kgCO$_{2}$e per chip for manufacturing emissions increased 1.8x from the v4i to v6e. Based on the breakdown of manufacturing GHG emissions, a larger TPU ASIC die---along with larger High Bandwidth Memory (HBM) modules packaged alongside the TPU chip (see Figure~\ref{fig:hardware-xsection-schematic} below)---and growing DRAM capacity in the host compute tray are the two most important emission drivers. Printed circuit board assembly (PCBA) and thermal are the two other important commodities, but increases in these two areas are smaller.

Among the three generations, manufacturing emissions for v4i and v5e are very close, but emissions for the latest generation (v6e) nearly doubled. This increase is mainly due to a threefold increase in host CPU host DRAM capacity: the v5e host has a memory capacity of 512 GB, whereas the v6e host has 1,536 GB DRAM. The large memory capacity in v6e is designed to support large language model training applications.

For the two powerful TPU platforms, TPUs (including ASICs and HBM) drive emissions increase over generations. Compared to v4, manufacturing emissions of v5p increased by 55\% (Figure \ref{fig:manufactuing-emissions}). This growth is mainly due to an 80\% increase in TPU manufacturing emissions. A breakdown of TPU emissions reveals that increases in HBM (95\%) are larger than the ASIC (70\%), as HBM capacity in v5p (96GB) is 3x larger than that in v4 (32GB). It is worth noting that the increase in HBM manufacturing GHG is lower than capacity growth because v5p is using a new generation of HBM that has higher bit density. The underlying wafer-level process flows yield lower carbon emissions when normalized by the functional unit of produced memory capacity. 

The CPU and IO emissions are noticeably higher for v5p versus v4, because the CPU used in v5p host has a much larger die size area than the CPU used in v4 host tray. The NIC card used in the v5p host server also has 3x higher GHG emissions due to more advanced chips and larger on-board DRAM capacity.

\subsection{Result 4: TPU emissions trends for two specific training tasks}\label{result:4}
This section provides a granular analysis of the carbon emissions associated with running identical example workloads on two TPU generations: v6e and v5e. This micro-level comparison offers insights into the CO$_{2}$e of jobs executed on different TPU platforms. 

The goal is to understand how improvements in theoretical carbon efficiency translate in real-world conditions, using specific examples. To ensure a meaningful comparison, we selected two very popular, representative, medium-sized training jobs from a broader set of models used to measure TPU performance:
\begin{itemize}
    \item Reinforcement Learning Human Feedback (RLHF).
    \item Supervised Fine-Tuning (SFT).
\end{itemize}
\textcolor{black}{These jobs were chosen due to their reasonable size and duration, which is especially important because our full-system energy measurements are available with five minute intervals. Some inference tasks and several MLPerf benchmarks had too short of a job duration to provide an useful and stable measure of full-system power consumption, necessary for an accurate view of carbon efficiency.}

Each workload was run multiple times on one v5e or v6e pod (32 machines). For each run, we measured the \textit{step time}, which represents the average time to complete one unit of work defined by the workload. To compare emissions, we use the \textit{emissions per step} metric,  $\textrm{Emissions}_\textrm{step}$, calculated as the sum of Operational $\textrm{Emissions}_\textrm{step}$ and $\textrm{Embodied Emissions}_\textrm{step}$, where:
\begin{equation*}
   \begin{aligned}
        \textrm{Operational Emissions}_\textrm{step} &=  (\textrm{measured on-duty machine power}) \\ 
        & \times (\textrm{step time}) \\ &\times (\textrm{operational carbon intensity}) \\
            \textrm{Embodied Emissions}_\textrm{step} &= (\textrm{embodied emissions rate}) \\
            &\times (\textrm{step time}). 
    \end{aligned}
\end{equation*}

The operational carbon intensity uses MB accounting from Section \ref{result: lca} and the embodied emissions rate comes from Section \ref{result:3} with an assumed 6-year lifetime.
\begin{table}[h]
    \caption{Average step time, power, CCI, and CO${_2}$e (MB) per step for both workloads on v5e and v6e.}
    \centering
    \begin{tabular}{| p{4.1 cm} | cc | cc |}
        \hline
         Workload & \multicolumn{2}{c |}{RLHF} & \multicolumn{2}{c |}{SFT} \\
         TPU &  v5e & v6e  & v5e & v6e\\
          \hline
         Step time (seconds) & 0.70 & 0.24 & 5.67 & 2.31 \\
         Average machine power (W)  &  1386 & 2589  & 1728 & 3156  \\
         \textcolor{black}{CCI (gCO$_{2}$e/10$^{18}$ FLOPs)}  & \textcolor{black}{309.9} & \textcolor{black}{183.8}  & \textcolor{black}{249.4} & \textcolor{black}{142.0}  \\
         $\textrm{Emissions}_\textrm{step}$  (gCO$_{2}$e)  & 0.044 & 0.027  & 0.418 & 0.307   \\
         \multicolumn{1}{| r |}{($\textrm{Operational Emissions}_\textrm{step}$)}  & 0.033 & 0.021  & 0.335 &  0.250  \\
         \multicolumn{1}{| r |}{($\textrm{Embodied Emissions}_\textrm{step}$)}
         & 0.010 &  0.006  & 0.082 & 0.057  \\ 

         \hline
     \end{tabular}
    \label{tab:carbon_per_step}
\end{table}

Table~\ref{tab:carbon_per_step} provides a detailed breakdown of the average step time, machine power, and emissions per step across all runs for both workloads on v5e and v6e. It also shows the CCI while the workload was running. As the data indicates, operational emissions dominate the total carbon footprint representing more than 70\% of total emissions in every case. For both models, moving from v5e to v6e substantially reduces total emissions per step. The observed carbon reduction from v5e to v6e is primarily driven by a decrease in step time, which is partially offset by the higher average on-duty machine power of v6e compared to v5e.

\textcolor{black}{
A few items make this result an interesting supplement to Result 2. First, the improvement for v6e's CCI over v5e's CCI is 41\% and 43\%, for the RLHF and SFT benchmarks, respectively. That's directionally similar to the Result 2 but smaller (2x improvement vs 3x improvement in the fleetwide result). That could indicate, for example, that some benefits in terms of improved CCI in new hardware could come from reduction in idle power. Or, it could be attributable to the small sample of just two benchmark workloads considered here. Future work could expand this analysis to a wider range of workloads to better explain the drivers of CCI gains.} 

\textcolor{black}{
Second, the relative improvement in directly measured emissions of the workload (i.e.  $\textrm{Emissions}_\textrm{step}$) are 39\% and 27\% for the RLHF and SFT benchmarks, respectively. So, the improvement in CCI and directly measured emissions is nearly identical for RLHF (39\% vs 41\%), but smaller for SFT (27\% vs 43\%).}

\textcolor{black}{We think this result supports the general claim that CCI is a good tool for approximating the carbon emissions of workload while highlighting that it is an estimate that does not perfectly capture the energy requirements and carbon emissions of a specific workload. Future work could help understand these differences.   
}

\section{LCA versus Corporate Inventories}
\label{appendix:EIR}
Our LCA found that the operational (electricity-related) emissions dominate the lifetime emissions of AI hardware (see Section~\ref{result: lca}). For the platforms we considered, approximately 25\% of AI hardware life-cycle emissions come from manufacturing, using the market-based operational electricity emissions. 

It would seem one could use a company's annual corporate emissions inventory to estimate the embodied versus operational emissions of AI hardware~\citep{Gupta2021}. However, in Google’s Environmental Report~\citep{Google2024}, about 32\% of Google's 2023 corporate emissions are attributable to hardware manufacturing of Tier 1 and beyond suppliers (Capital Goods and Other categories). 

Surprisingly, this gap is expected due to the \textit{differing calculations} of this LCA and Google's corporate inventory. This gap comes from these key distinctions between the two approaches:
\begin{enumerate}
    \item \textbf{Accounting Methodology}: The LCA approach follows ISO 14040 and 14044 standards that consider emissions associated with a well-defined functional unit. These standards allow for the amortization of emissions over its lifetime. However, Google's corporate inventory~\citep{Google2024} follows the GHG Protocol: Corporate Standard that \textit{does not allow for amortization of Capital Good emissions across hardware lifetimes.} Thus, emissions for a given hardware purchase will be disclosed fully in the first year of use, while only $\sim$1/6th of the hardware lifetime and $\sim$1/20th of the data center lifetime occurs in the first year. This accounting approach is the largest difference between emissions from  the LCA and corporate inventories. 
    \item \textbf{Other Emissions Sources}: The LCA considers only the functional unit of the AI accelerator and host machine. The corporate inventory of Google includes a wider inventory boundary that includes emissions sources unrelated to AI, including emissions outside the data center and emissions associated with manufacture of consumer devices. Some of these emissions are unrelated to AI.
\end{enumerate}

Due to the different methodologies and the continuous growth of Google's data centers, the annual reported hardware emissions in Google's corporate inventory are driven by first-year hardware purchases and the rate of data center growth, and therefore do not directly correspond to a lifecycle view of the machines deployed by Google.

\section{Related work}
\label{section:related}
The existing literature provides targeted assessment of the environmental impact of certain AI workloads, but often lacks a comprehensive and consistent view of these impacts. Here we provide the context of many of these studies following the cradle-to-grave life-cycle that Figure~\ref{fig:lifecycle-diagram} above defines.

\textbf{AI accelerator hardware manufacturing emissions} and life-cycle assessment studies are relatively limited in the literature. \citet{Kuo2022}~show that improvements in semiconductor manufacturing processes and node size have led to reduced life-cycle manufacturing emissions of DRAM; DRAM is an important component of AI accelerators,  as DRAM memory is one of key drivers of the overall AI hardware footprint in our results. \citet{Luccioni2022}~estimate emissions associated with server and GPU hardware of the open-access BLOOM language model. For the embodied emissions of AI, they use a placeholder for the GPU hardware, which the original source calls an ``arbitrary carbon footprint value'' of 150 kg ~\citep{Davy2021}.  Our TPU results are 1.5x--4x larger at 208--585 kg (TPU M+T in Table ~\ref{tab:specs}). Extrapolating from a 2019 Apple Mac Pro LCA ~\citep{Apple2019}, \citet{Wu2022} use 1808 kg for a whole GPU server; TPU machines average 1.7x larger at 2300--4672 kg.
~\citet{Gupta2022} estimate the embodied carbon of logic chips, DRAMs, SSDs, and hard disks.
\citet{Wang2024}~sketch the design of a carbon-efficient compute server using low-carbon components; they address standard servers, not AI hardware. 

While there's no full LCA for AI hardware,~\citet{Ji2024} reviews 6 known public LCAs of server computers. Their embodied emissions vary by 40x, from 423–15,593 kgCO$_{2}$e, averaging $\sim$3900 $\pm \sim$5900 kg. \citet{Wu2022} use 909 kg of embodied emissions for a CPU server host extrapolated from an Apple Mac Pro LCA. The embodied carbon for our machines average $\sim$3300 $\pm \sim$900 kgg, including the CPUs and TPUs. (The CPUs without TPUs average $\sim$1200 $\pm \sim$500 kgCO$_{2}$e.) For the compute servers in~\citep{Ji2024}, using a 4 or 5 year lifetime, operational emissions are 70\%–90\% of total LCA emissions for 2 Dell servers, 66\%–94\% for 2 HP servers, and 39\%–97\% for 2 Lenovo servers. For our servers, they are 70-90\% of emissions over a 6 year lifetime (using MB at the low-end, LB at the high end).

\textbf{AI accelerator hardware retirement emissions} associated with ultimate disposal, recycling, and cascaded use of data center hardware have limited study for AI hardware in the literature. This paper shows its size relative to other AI hardware lifecycle emissions to help studies of AI accelerator hardware retirement.

\textbf{AI accelerator operational emissions}, on the other hand,  have been a substantial focus of research. Individual papers often focus specifically on AI model training (development) or AI serving. Emissions associated with energy usage of AI model training have been the most studied emissions source in the literature. 

Early studies~\citep{Strubell2019,Patterson2021} focused primarily on estimating the energy-related emissions of training workloads. These studies found results to be very sensitive to input assumptions, such as the \textcolor{black}{electricity emission factor} and the sampling window of the training energy. A generational meta-analysis by \citet{Luccioni2023} found that training emissions per model have increased by a factor of approximately 100 from 2012 to 2023. Informed by previous studies, and using best-practices to improve training carbon efficiency, \citet{Touvron2023}~published an estimate for Meta’s production of Llama 2 emissions at 300 tonnes of CO$_{2}$e. 

Serving emissions associated with the inference and querying of AI models have garnered interest as inference begins to scale with consumer usage of AI products~\citep{patterson2024energy}. \citet{Luccioni2022}~first estimated weekly inference emissions associated with the BLOOM model. \citet{Patterson2022}~estimate 60\% of machine learning energy use at Google is attributable to inference, and \citet{Wu2022}~find some Meta use cases (namely LLMs) generate two thirds of their operational emissions during inference. These results illustrate the importance of taking a holistic approach to AI emissions measurement, since inference emissions are likely to increase their share of the total emissions of AI products as their adoption continues. 

\citet{Gupta2021} uses Google and Meta corporate emissions inventories to observe that data center emissions are shifting from operations to hardware design, manufacturing, and construction, e.g., they were half of Facebook’s 2019 Scope 3 emissions. Using LCAs, our results differ. Section ~\ref{appendix:EIR} explains the differences in accounting of ERs and LCAs. In particular, emissions for new data centers and new hardware must be fully accounted for in a single year for an ER rather than amortized over their lifetimes in an LCA.  \textcolor{black}{In addition, Section \ref{result: lca} points out the market-based emissions accounting practices are not identical across companies.}

\textbf{AI hardware \textcolor{black}{carbon intensity} metrics} are necessary to normalize emissions relative to their performance or value-creation. There is not yet consensus on a consistent \textcolor{black}{carbon intensity} metric, but typically CO$_{2}$e or energy / performance is deemed suitable~\citep{Jouppi2023,Vahdat2024}. Our proposal for the metric is CO$_{2}$e/ExaFLOP, labelled \textit{compute carbon intensity (CCI)}.

Some use machine TDP to infer hardware energy consumption (e.g., \citep{bouza2023estimate,lannelongue2021green,trebaol2020cumulator}), but we found that the ratio of TDP to actual average power varies from 2X--6X, so TDP dramatically overestimates the real result for TPUs.

\textcolor{black}{\textbf{Software optimizations} led to even larger improvements than hardware gains. For example, \citet{hernandez2020measuring} found algorithmic advances yielded a 44x performance improvement in 7 years and \citet{ho2024algorithmic} calculated that over 11 years they halve computation demands every 8 months. Similarly, the cost per token of commercial inference services has dropped by 10X in the past 2 years. Thus, it is likely that software optimizations will further reduce the carbon footprint per token in future years.}

\section{Conclusion}
We believe this is the first published study of a cradle-to-grave analysis of the carbon footprint of AI hardware, including the first publication of manufacturing emissions of an AI accelerator. To complete our evaluations, we needed a new \textcolor{black}{carbon intensity} metric related to CO$_{2}$e/goodput called \textit{compute carbon intensity} (CCI), measured as gCO$_{2}$e/ExaFLOP. We believe this metric will be useful for many in evaluating AI hardware emissions comprehensively, and helpful in estimating the carbon footprint of their workloads. Ideally, CCI would join performance/TCO~\citep{Jouppi2021} as a design target and the detailed appendices will act as a tutorial, road map, and inspiration that encourages more engineers to perform LCAs.

\textcolor{black}{The CCI metric can help clarify the multiplicative impacts of hardware and software improvements.  Hardware improvements reduce CCI: g CO$_{2}$e/ExaFLOP. Software efficiency improvements reduce the number of computations (FLOPs) required relative to the model quality. Improving each by 3x would result in a 9x reduction in carbon emissions for the same model quality.} 

Our evaluation led to the following observations:
\begin{enumerate}
    \item More advanced TPU chips and increasing memory demand drive embodied emissions---representing more than half of all embodied emissions with memory alone more than a third---yet CCI from manufacturing still declines each generation, suggesting performance gains via more efficient hardware design outweigh increases in manufacturing emissions.
    \item Operational emissions dominate AI hardware lifetime at 70\% to 90\% of total emissions, with manufacturing emissions under 25\% and data center construction emissions under 5\%.
    \item When evaluating AI benchmark workloads, workload emissions decline per TPU generation. For example, gCO$_{2}$e per training step dropped 27\% to 39\% over one TPU generation when training the same model.
    \item CCI as measured across Google's fleet declines with each TPU generation, delivering a 3x improvement in CCI from TPU v4i to TPU v6e (Trillium).

\end{enumerate}

\textcolor{black}{
Our data shows a 3x hardware improvement in just two generations of AI hardware. This doesn't account for Google’s continued carbon-free energy (CFE) procurement, which drives an additional reduction in the lifetime carbon emissions of TPUs. Coupled with the hardware improvements, a v6e TPU deployed in a 90\% CFE data center could hypothetically achieve a 10x improvement in CCI relative to a TPU v4i. Similarly, the 3x improvement doesn't factor in software efficiency gains, like Gemini 2.0 Flash outperforming Gemini 1.5 Pro at twice the speed~\citep{googleGeminiFlash} (Section \ref{section:related}). If software advancements halve the required FLOPs for the same model quality, the overall carbon emissions reduction for that AI model could reach an impressive 20x. Combined advancements in hardware design, hardware energy efficiency, CFE procurement, and software efficiency offer multidimensional paths to reducing the carbon footprint of AI.}

\begin{acks}
Thank you to our colleagues at  who helped shape and improve this research and the resulting paper. Thank you especially to Savannah Goodman, Maud Texier, Dani Ton, and Thomas Olavson, who helped shape the direction of the research and provided critical feedback. Thank you to David Culler, Urs H{\"o}lzle, Ben Gomes, Norman Jouppi, Benjamin Lee, David Lo, Xiaoyu Ma, Devon Swezey, and Chrissy Patterson for feedback on the draft.
\end{acks}

\appendix
\section{Steps of a Life-Cycle Analysis (LCA)}
\label{appendix:steps}
To assess supply chain manufacturing GHG emissions, we performed machine-level \textit{cradle-to-gate} (i.e., from start to delivery) \textit{Life-Cycle Assessments} (LCA), consistent with international LCA standards (ISO 14040 and 14044), applying industry best practices. 

For readers unfamiliar with an LCA, we believe the  appendices can be a tutorial, as they detail all the steps to perform an LCA. After defining the boundaries between the development and use stages, this appendix shows how to properly collect emissions data over the lifetime of computer equipment, including data centers.  Appendix \ref{appendix:TPUv5} then gives a concrete example of an LCA for TPU v5e.
Appendix~\ref{appendix:24/7} introduces 24/7 emissions and Appendix~\ref{appendix:Methodology} describes efforts to standardize carbon footprint assessments for electronics.

\subsection{Inventory boundary and exclusions} \label{appendix:exclusions}
For LCA modeling, the system boundary covers all stages along the supply chain and usage phases, including manufacturing, operations in data centers, and end-of-life disposal. On the other hand, since corporate emissions are typically reported following the GHG Protocol~\citep{GreenhouseGasProtocol2004}, we define the emission inventory boundary of this study along the operational control consolidation approach and divided life-cycle emissions into three scopes. Operational control includes activities where the developer has control to implement operational changes. In order to provide a comprehensive view of AI emissions sources, we follow a proposed emission inventory boundary that includes all cradle-to-grave emissions of an AI product life-cycle including:

\noindent Scope 1: Operational direct emissions
    \begin{itemize}
        \item Fossil fuels combusted on-site, such as diesel for backup power, natural gas for heating, and fuels fleet vehicles use.
        \item Fugitive emissions from data center HVAC system coolants.
    \end{itemize}
\noindent Scope 2: Operational electricity emissions\footnote{Note: Despite the distinction between Developing and Serving emissions in Figure~\ref{fig:lifecycle-diagram}, this paper focuses on AI hardware in aggregate across Google's fleet and therefore presents a combined operational emissions measurement in the results section.}
    \begin{itemize}
        \item Direct energy measurements of deployed machines, which are engaged in all parts of the AI software stack, including:
        \begin{itemize}
            \item AI model development, experimentation, and training.
            \item AI model bulk inference and fine tuning for foundational model application to specific product use cases.
            \item AI model inference for end-user queries.
        \end{itemize}
        \item Data center overheads, including for machine cooling. 
    \end{itemize}
\noindent Scope 3: Embodied hardware and data center emissions
    \begin{itemize}
        \item Data center hardware manufacturing following a cradle-gate-approach,
including raw material extraction, refinement, components and sub-assembly manufacturing, and final server configuration.
        \item Data center hardware transportation, including upstream shipping of materials and parts to manufacturing locations, and shipping of assembled servers from contract manufacturers to Google’s data centers.
        \item Data center hardware reverse logistics, including potential reuse, material recovery and end-of-life treatment.
        \item Data center building construction, including manufacture of construction materials.
    \end{itemize}
Notably, this definition of operational control excludes, which are small or outside of AI developers' operational control:
\begin{itemize}
    \item \textcolor{black}{Upstream emissions from
    purchased electricity (e.g. emissions from coal plant construction), as well as emissions from transmission and distribution losses.}
    \item Small equipment deployments (i.e., edge nodes) at internet service providers' partners.
    \item Consumer device operational emissions (e.g., laptop usage to query an AI product) and consumer device end-of-life.
\end{itemize}

\subsection{Manufacturing emissions} \label{appendix:Emissions}
At the core of our semiconductor assessments we use a combination of advanced life-cycle inventories for front-end and wafer-level processes based on IMEC’s virtual fab. We customized these analyses to Google’s chip designs incorporated into our platforms, and proprietary life-cycle inventories to account for the remaining back-end chip and panel-level assembly. The virtual fab in IMEC.netzero ~\citep{IMEC2024} utilizes a detailed, bottom-up approach to assess environmental impacts. It starts by analyzing process flows, recipes, and tool data to create a virtual model of a high-volume semiconductor manufacturing facility. This model incorporates a variety of logic and memory technologies, i.e., DRAM and NAND, across different production nodes (both current and upcoming). Moreover, advanced packaging life-cycle inventories allow for the configuration of HBM stacks and the interconnections between ASIC and memory components on silicon interposer. The accuracy of this model relies on detailed fab models that realistically simulate process flows and individual steps. Impact analysis is performed after customizing manufacturing-related die-related parameters, such as technology nodes in multi-chip-modules, die sizes, yield rates, custom electricity mix and fluorinated GHG abatement ratios. 

\begin{figure}[t]
	\centering
	\includegraphics[width=0.45\textwidth]{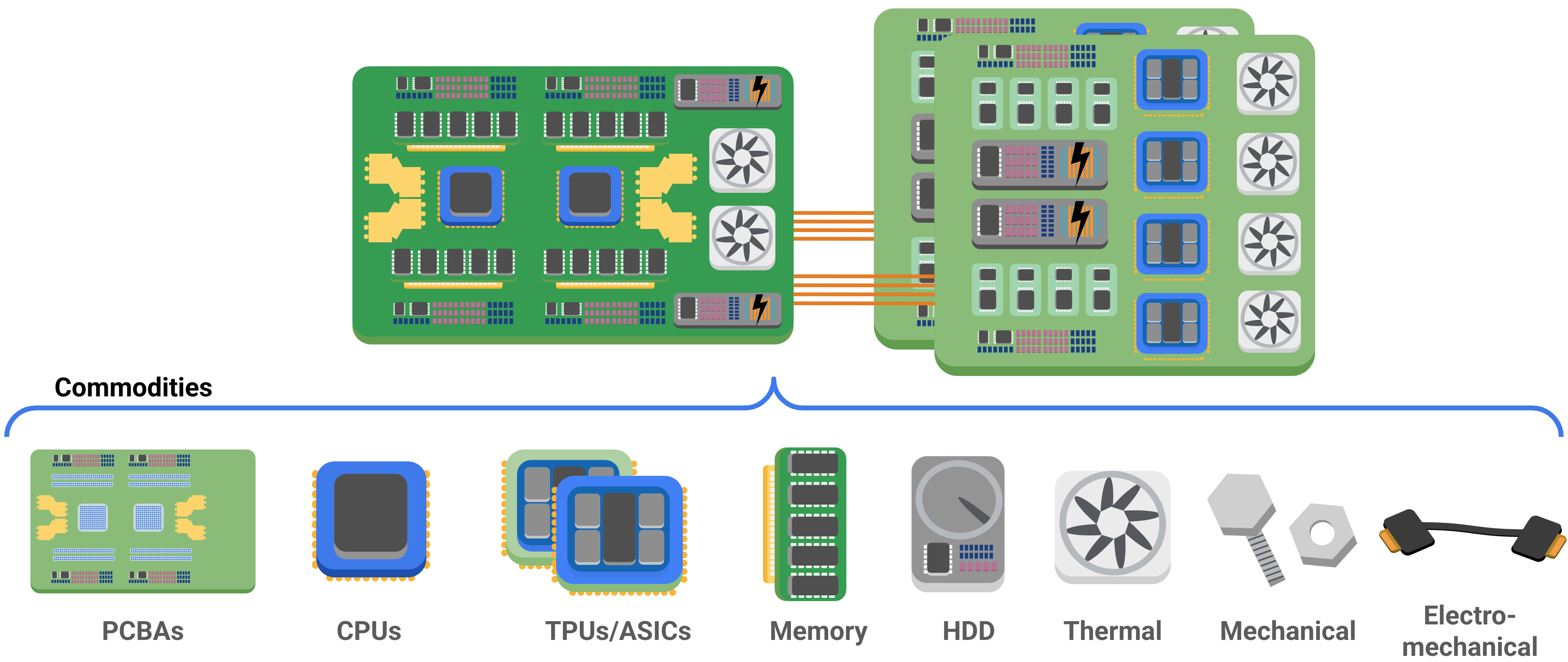}
	\caption{Typical hardware components in a server.}
	\label{fig:hardware-diagram}
\end{figure}

This analysis allows for a thorough understanding of the environmental footprint associated with semiconductor manufacturing, and provides insights capable of reflecting emissions at the process level and to tracing emissions down to their root causes. For other emission hotspots, i.e., our thermal management solutions, we collect supplier data to account for material types, scrap rates, and metal forming electricity consumption and grid mix.

\subsubsection{Life-cycle inventory data}
The \textit{Life-Cycle Inventory} (LCI) data used for AI machine LCAs is a combination of primary and secondary (or industry average) datasets---see Table~\ref{tab:lca-data}---collected through a combination of various methods and data sources, including bill of materials (BOMs), GCM engagement, and supplier-provided primary data. When necessary, detailed teardowns are conducted on platform trays to confirm completeness of components captured through above-mentioned methods and fill data gaps on part dimensions and materials. 

 The product system scope follows a bottom-up approach considering all physical components that make up individual trays. Figure~\ref{fig:hardware-diagram} shows that for a typical tray, major components include the processors (e.g., CPU or TPU), memory (e.g., DRAM, SSD), printed circuit boards assemblies (PCBAs), thermal management solutions (e.g., heatsink, cold plate), mechanical components (e.g., base tray enclosure), and electromechanical components. For the selected TPUs, the full list of hardware components is mapped to the internal LCA commodity-level data repository. 

We build on comprehensive data repositories that are independently third-party reviewed, by deploying a combination of Sphera Professional Databases~\citep{Sphera2024} and ecoinvent ~\citep{Frischknecht2005}. Both data sources are the most established among LCA practitioners in the \textit{Information and Communication Technology} (ICT) sector and are frequently updated to remain accurate, relevant, and comprehensive. Our foreground models allow for the necessary configurability of the LCIs to ingest key primary data points relevant to technologies in scope and supplier-specific activity. While we customize chip production using IMEC’s virtual fab, commodities such as PCBAs, mechanical enclosures, and thermal management, are configured through proprietary, parameterized models in LCA for Experts.

\begin{table}[t]
	\caption{Data source by component type. TPU, CPU, memory ICs, and modules are accounted for separately.}
	\label{tab:lca-data}
	\centering
    \begin{tabular}{p{1.2cm} p{3.5cm} p{3cm}}
        Commodity & Primary data & Secondary Data \\
        \midrule
        TPU, CPU, Memory ICs and Modules & Electronics package layout, interconnects, die sizes, technology nodes, front-end process abatement, yields, region of production & IMEC.netzero virtual fab for wafer-level processes, carbon intensity of energy and materials used in back-end manufacturing \\
        \midrule
        PCBA & Substrate layer count, dimensions, panel layout, metallization and full BoM of mounted devices & Carbon intensity of substrate, chips, and onboard mechanicals, connectors and wires \\
        \midrule
        Thermal Management & Heatsink and cold plates mechanical designs, materials, scrap rates and electricity consumption in metal forming processes and final assembly, region of production. Fan dimensions and weight. & Carbon intensity of upstream materials, off-the-shelf components and energy consumed in metal forming and final assembly \\
        \midrule
        Mechanical & Material types, dimensions, weights, scrap rates, electricity emissions factors & Carbon intensity of upstream metal sheet production and energy used for metal forming \\
        \midrule
    \end{tabular}
\end{table}

\subsubsection{Life-cycle impact assessment}
Once an LCI completes, we assess the cradle-to-gate environmental impact of an AI machine by characterizing its LCI using \textit{life-cycle impact assessment} (LCIA) ~\citep{klopffer2014life}. This process quantifies energy and material flows included in the system boundary, and measures the environmental outcome using a common metric. Since our focus is carbon emissions, the metric for climate change impact is typically expressed as carbon intensity per functional unit, such as kg CO$_{2}$e per kg of metal used for base tray, or kg CO$_{2}$e per kWh of electricity used for chip production. GHG beyond CO2 (e.g., N2O, CH4, SF6, etc.) are converted to CO$_{2}$e equivalents using global warming potential (GWP) ~\citep{Houghton1990} as the environmental impact indicator of choice in this study. Because GHG emissions include not only CO2, we need a metric to quantify the relative contributions of different substances. The GWP is commonly used to convert climate change impact of non-CO2 emissions to the CO$_{2}$e equivalent. We use GWP for a one hundred year timeframe (GWP100) from the 5th assessment report (AR5) of the Intergovernmental Panel on Climate Change (IPCC) ~\citep{Myhre2013}.

\subsubsection{Transportation emissions}
The hardware transportation GHG emissions for AI machines include transportation emissions associated with shipping AI trays from manufacturing sites to data centers. For the AI accelerator tray, transportation GHG emissions include three major segments: 1) ground shipping from the manufacturing site in Asia to a nearby departing airport, 2) shipping to a Google hub in the US, and 3) ground transportation from the hub to individual data centers. For host trays, transportation emissions are modeled as a combination of sub-assemblies. For mechanical components, transoceanic travel is modeled as ocean shipping. For the rest (chips, PCBA), transoceanic travel is modeled as air travel. Once they arrive in the US, ground transportation emissions from hubs to individual data centers are added.  Emissions reflect weight of both the tray itself and the packaging portion. 

Appendix \ref{appendix:TPUv5} further illustrates the LCA processes of this appendix, using TPU v5e as an example. 

\subsection{Operational (Development and Serving) electricity emissions}
\label{Appendix:Operational}
This section describes the methodology to estimate the energy consumption and electricity-related emissions of AI hardware in Google's fleet. In Section~\ref{section:definition} we estimated the \textcolor{black}{carbon intensity} of electricity consumption in the data centers where AI hardware is deployed. To estimate the electricity-related emissions from a specific machine (e.g., TPU v5p), we also need to collect power / energy measurements from all machines of that type deployed in Google data centers.
We leverage existing hardware and software systems built by Google to allow for measurement and estimation ~\citep{Schneider2024}. 

For each AI hardware device in this study, we measured power consumption from each tray via its \textit{power supply unit} (PSU), which includes an estimate of 4\% overhead for rectifier losses. We use these measurements from the PSU of each machine tray to collect average AI machine power data. To get the total energy measured for a machine, which may include multiple trays, we sum the total power measured across all tray PSUs, plus estimated rectifier losses. Thus, the number of measurements is a function of the number of trays per machine: e.g., two for v5p (motherboard and one TPU tray) and three for v5e (motherboard and two TPU trays). In each case, the TPU tray itself includes a printed circuit board with four TPU chips; see Figure~\ref{fig:hardware-diagram} above for an example. 

Tray power data is collected from a central Google database, where measurements are collected every five minutes. For PSU power data, the reading is the average power consumption measured at the PSU sensor for the past five minutes, so it represents average power. In some cases, a small portion of machines may have missing power readings for trays (e.g. due to sensor failures), and these observations are excluded from analysis.

To get the average energy consumption per day, we take the total energy consumption of all machines of a particular machine type over a representative month (e.g., October 2024), and then divide by the number of machine-days in that sample. To get the average lifetime operational energy consumption for a machine type, we then multiply this value by the assumed six year machine lifetime and the average Google data center \textit{power usage effectiveness} (PUE). Google’s 2024 average PUE is 1.10 ~\citep{Google2024}. 

\subsection{End-of-Life emissions}
By applying Google’s Zero Waste to Landfill strategy~\citep{Google2024}, cascaded use (through a combination of device lifetime extension and end-of-life material recovery) leads to net carbon credits if deployed in closed loops. We estimate through LCA, that the latter has the potential to offset embodied emissions by up to 4\%, applying the avoided burden approach to value of scrap material. Bulk metal scrap is collected from uniform aluminum, steel and copper alloys found in macroscopic components (e.g. thermal management, interconnects and enclosures), while precious metals, incl. Ag, Au, Pd, and Pt are recovered as by-products of the copper electrolysis applied to shredded PCBAs. However, for the purpose of this paper, we take a conservative assumption and do not claim credit for the avoided burdens as the absolute amount will depend on the individual product’s second life and end-of-life treatment route.

\begin{figure}[b]
	\centering
	\includegraphics[width=0.25\textwidth]{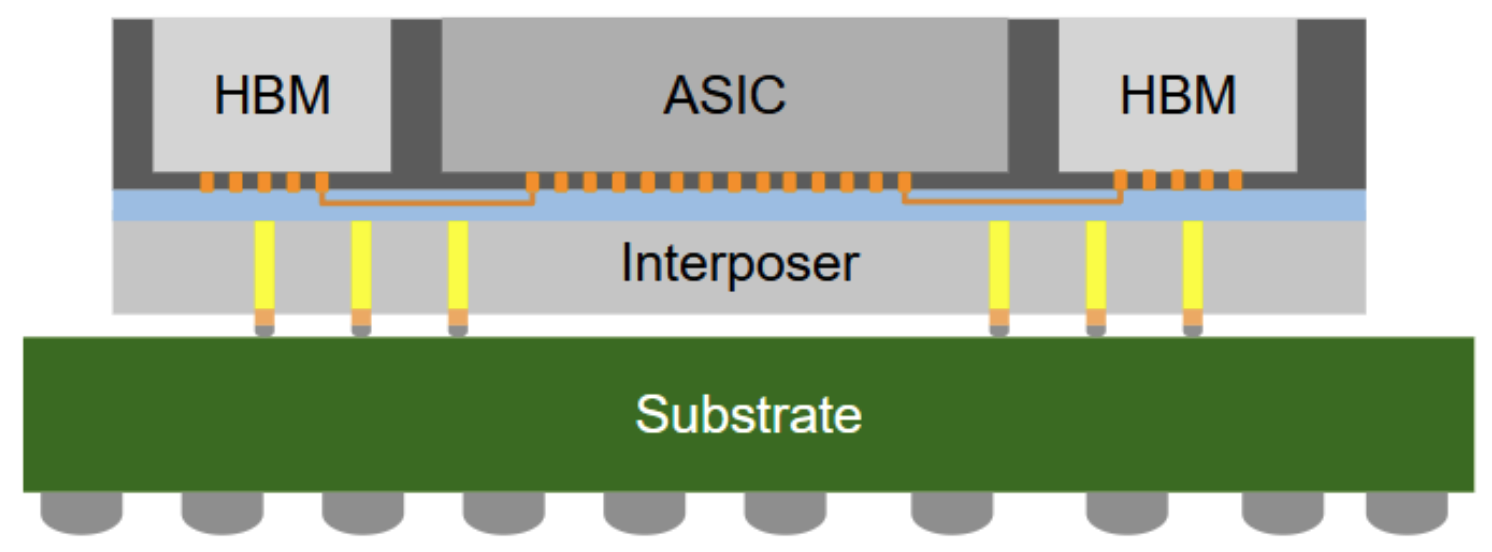}
	\caption{A generic schematic cross-sectional sketch of a TPU 2.5D package. Size is not scaled to actual hardware.}
	\label{fig:hardware-xsection-schematic}
\end{figure}

\subsection{Data center construction and Scope 1 emissions}
\label{appendix DC}
Two additional contributors of AI life cycle emissions are the construction of data centers that house AI hardware and Scope 1 emissions from local backup generation for data center emergencies. Since these emissions are very small in the life-cycle of AI hardware (<5\%), we reuse a simple existing approach from Google's Cloud Carbon Reporting~\citep{Cloud2024} to roughly estimate their emissions.

\subsubsection{Embodied emissions of data center facilities}
This emission source encompasses the embodied emissions of data center construction materials and the emissions associated with the construction itself. This includes site infrastructure such as coolant systems and power systems. Using life-cycle analysis, Google has established a data center construction emissions footprint, accounting for the size of new data center additions. This scaled footprint is then amortized over 20-years.  

\subsubsection{Scope 1 - Fossil fuels combusted onsite \& fugitive emissions from onsite HVAC (heating, ventilation, air conditioning)}
These emissions include all data center onsite fuels use---e.g., for backup power, water and space heating, and transportation (fleet vehicles)---and fugitive emissions from HVAC leakage. Google calculates the resulting carbon footprint in its emission reports; we apply a fraction proportional to each TPU's share of overall energy consumption.

For both categories, total annual emissions are allocated across all data center machines and workloads relative to their fraction of total annual data center energy consumption.

\section{Example: Embodied Emissions of TPU v5e}
\label{appendix:TPUv5}
\subsection{Manufacturing and transportation emissions}
Taking the TPU v5e machine as an example, the LCA process starts with compiling bills of materials for the accelerator tray and host tray. Based on the components on each tray, a detailed LCI database is constructed and used for LCA modeling. For large integrated circuits like TPUs, we use the IMEC.netzero tool and technology-node specific data sets that were developed with the industry-joint IMEC SSTS program. This new tool is a big improvement in integrated circuit manufacturing modeling capabilities by allowing us to:
\begin{enumerate}
    \item Model advanced tech nodes (e.g., 5 nm, 3 nm, 2 nm) incorporating materials and energy impact of advanced packaging technologies, e.g., through-silicon-vias (TSVs) in HBMs, and 2.5D stacking via silicon interposer 
    (see Figure~\ref{fig:hardware-xsection-schematic}).
    \item Apply supplier-specific customization of direct emissions abatement efficiency.
    \item Use country-specific electricity mix for each component, e.g., different countries fabricate HBMs and ASIC chips.
\end{enumerate}
For the rest of the hardware components, we use the \textit{LCA for Experts} software application and a combination of primary data and industry average data (see Table~\ref{tab:lca-data} above).

The LCA results suggest manufacturing GHG emissions add up to a GWP100 of 747 kgCO$_{2}$e per TPU v5e accelerator tray, and 782 kgCO$_{2}$e for the host tray. At the machine level (i.e., two accelerator trays per host tray), total manufacturing GHG emissions are estimated at 2,277 kgCO$_{2}$e/machine. Transportation would add another 471 kgCO$_{2}$e to machine-level scope 3 GHG emissions. For the host tray, the DRAM Dual In-Line Memory Modules and mainboard are the dominating GHG contributors (Figure~\ref{fig:manufactuing-emissions}). For the v5e tray, TPU chips and thermal management solutions (heatsinks, forced active cooling) contribute 62\% of tray-level manufacturing GHG. The remaining portion is attributed to the mainboard, tray enclosure, network interface card (NIC), and miscellaneous components.

\subsection{Operational emissions}
Following the methodology outlined in Appendix ~\ref{appendix:steps}, Table ~\ref{tab:specs} above presents the measured power for TPUs. Measurements of actual measured power in the fleet illustrate the importance of using measured data rather than machine thermal design power (TDP), which can be 2x to 3x higher. This result justifies the use of average hourly power in this study; the alternatives, TDP or a fixed fraction of TDP, are very coarse approximations. To account for data center overheads, we multiply by Google’s average data center PUE, 1.10~\citep{Google2024}.

\section{\textcolor{black}{Introduction to 24/7 CFE}}
\label{appendix:24/7}
\textcolor{black}{
A number of leading academic studies find that annually matched CFE procurement has less impact than hourly matching~\citep{xu2024system,riepin2024means,IEA2022}. One problem is the GHG Protocol Scope 2 Guidance~\citep{sotos2015ghg} allows companies to reduce their electricity emissions by matching annual electricity use with CFE purchases including from unbundled EACs (Section ~\ref{result: lca}) that may have little impact on reducing actual system emissions~\citep{bjorn2022renewable}. Companies that purchase solar or wind energy to match 100\% of their annual electricity consumption may still rely on carbon-emitting grid electricity for over 50\% of their demand~\citep{de2019100}. This discrepancy has become even more significant as the rapid progress of AI has increased the electricity demand in a number of regions where fossil generation still provides a significant share of electricity and where new clean power solutions are needed~\citep{IEA2024}. 
} 

\textcolor{black}{To try to address these inaccuracies, in 2019 Google developed another electricity accounting system---what we call \textit{24/7 Carbon-free Energy} (CFE)---that goes beyond the traditional Scope 2 rules to restrict both the location and time period where CFE purchases can be applied to reduce Scope 2 emissions. Using this method, a MWh of purchased CFE must be matched to a MWh of consumed electricity that is delivered to the same local grid boundary and in the same hour where consumption occurs. This more accurate accounting disallows CFE purchases made from different geographies and which are not matched with hourly consumption from reducing our electricity emissions. For this reason, the 24/7 emissions metric is higher than current MB emissions, while still reflecting emissions reductions relative to the LB metric.  The 24/7 Carbon-free Energy Compact was launched in 2021 by Sustainable Energy for All and UN-Energy to help organizations to achieve this goal and it has been signed by more than 160 companies, governments, and organizations~\citep{UN2021}.
}

\textcolor{black}{
To move closer toward our 24/7 CFE goal, Google typically signs long-term contracts, including \textit{Power Purchase Agreements} (PPAs). These agreements provide revenue certainty to CFE producers that enables them to finance and construct new generating assets. Google works to purchase clean energy from a variety of carbon-free sources as well as energy storage to increase the percentage of local and hourly matched CFE supplying our data centers and office campuses. 
}

\textcolor{black}{The percentage of CFE for a datacenter is reported ex-post, after load, production, and grid mix data are settled and made available to Google. With the current 24/7 CFE framework, when Google cannot get 100\% CFE from the grid plus its clean energy contracts in a given hour, the shortfall counts against the goal. 
}

\textcolor{black}{24/7 CFE accounting provides a more accurate inventory of a company’s electricity emissions and how effectively they are mitigated through purchases of CFE. It also enables companies to make procurement choices that bring us closer to a 24/7 CFE grid, faster. Reaching zero emissions on a 24/7 CFE basis (e.g. 100\% CFE) will require companies to consider how different CFE technologies can be combined to deliver an increasing amount of CFE around the clock. Many studies have shown that this technology diversity accelerates decarbonization of the broader electricity system, including by accelerating the commercialization of advanced CFE technologies~\citep{xu2024system,riepin2024means,IEA2022}.}

\textcolor{black}{For these reasons, we recommend that the AI community adopt these more accurate and impactful accounting standards for calculating Compute Carbon Intensity and for addressing the emissions associated with AI electricity consumption.}

\begin{figure}[b]
	\centering
	\includegraphics[width=.47\textwidth]{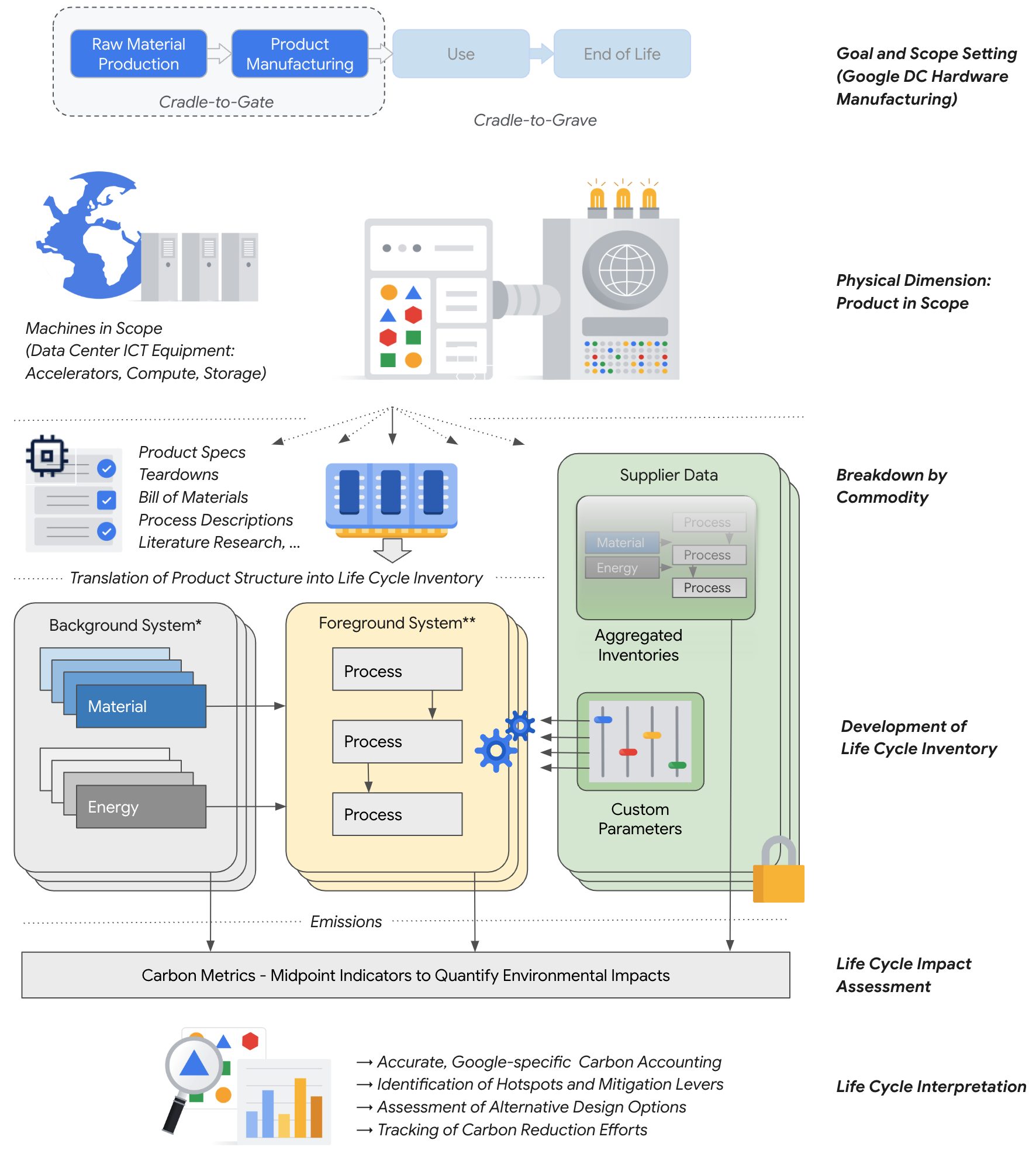}
	\caption{Schematic of the process flow for the manufacturing emissions methodology.}
	\label{fig:process-flow}
\end{figure}

\section{Initiatives and Process Flow of Manufacturing Emissions Methodology}
\label{appendix:Methodology}
Figure~\ref{fig:process-flow} gives an example of the process flow followed in this paper. 

To try to standardize product carbon footprint assessments for electronics, Google is actively collaborating with industry partners on unified methodologies for conducting LCAs for specific Information and Communication (ICT) devices. A key focus is the development of Product Category Rules (PCRs) that aim to streamline supplier data collection and ensure comparability across assessments, with the overarching goal of aligning with established standards like ISO 14040/44/67, the GHG Protocol, and EU PEF guidelines. We started with the impact category of climate change, planning to expand to additional impact categories in the midterm. 

While Google is already sponsoring the development of a PCR for computers and laptops ~\citep{spiliotopoulos2024methodological}, the initiation of a parallel industry-joint workstream specific to Data Center Equipment is underway via the Open Compute Project’s Sustainability project. This collaborative effort will involve documenting existing LCA practices and further developing configurable, parameterized model building blocks relevant to data center equipment, ensuring alignment between data developers. Additionally, the initiative will define sufficient data accuracy and coverage to enable the identification of hotspots, drive supply chain action, and facilitate carbon-aware product design decisions. A key outcome will be the establishment of a standardized approach to primary data reporting throughout the supply chain, contributing to a more harmonized and transparent approach to carbon footprint assessments in our industry. 

\section{On-Duty Machine Power for Benchmark Workloads} 
\label{appendix: benchmark_power}
Accurately determining on-duty machine power is crucial for computing the Scope 2 carbon emissions per step for Result 4 in Section \ref{result:4}. To achieve this accuracy, we collected machine power and duty cycle data for each machine at five-minute intervals throughout each workload run. 

\begin{figure}[t]
	\centering
	\includegraphics[width=.4\textwidth]{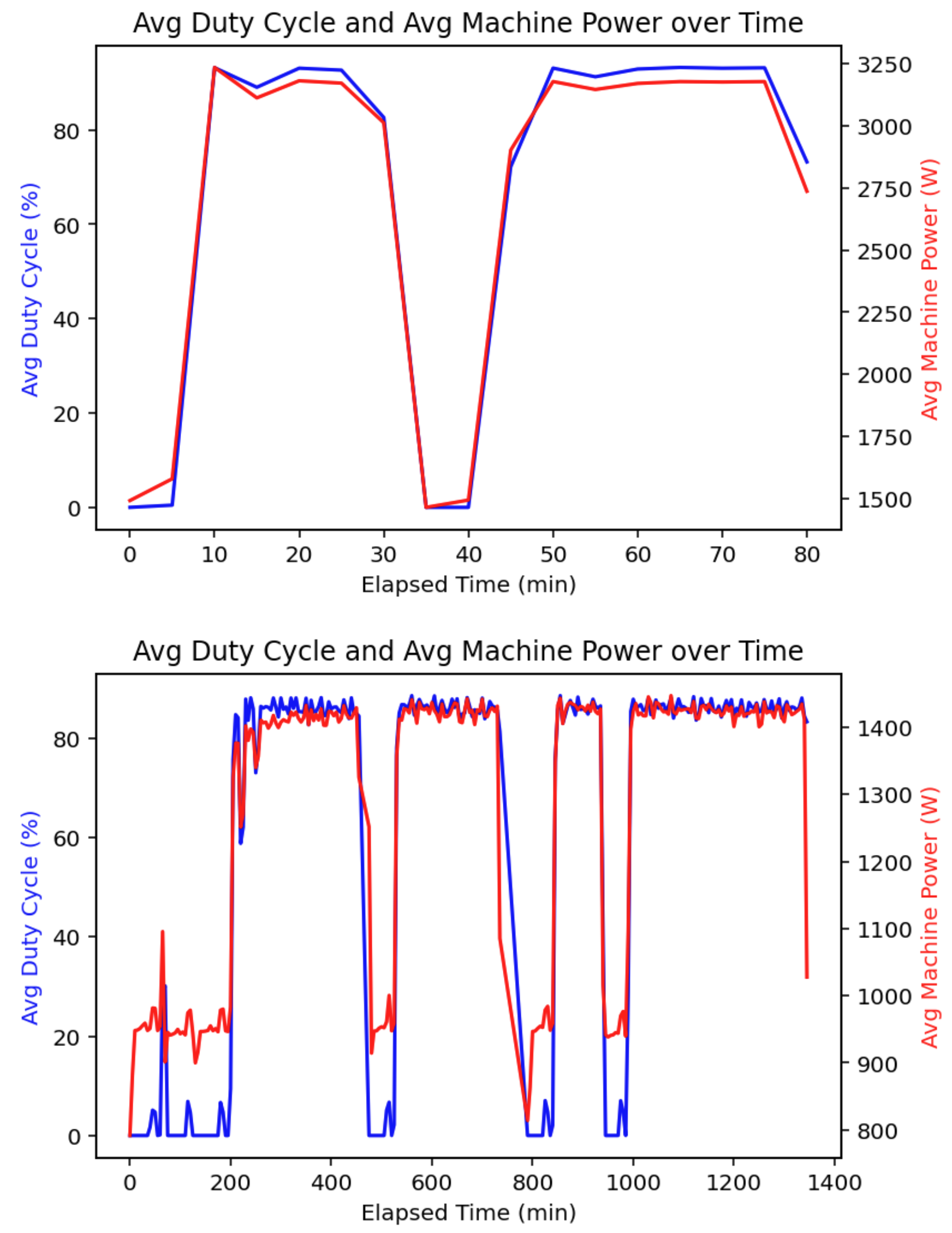}
	\caption{Duty Cycle and Power over time for RLHF on v5e (top) and for SFT on v6e (bottom).}
	\label{fig:duty-cycle-power}
\end{figure}

Figure \ref{fig:duty-cycle-power} displays this data for two representative runs. As illustrated, average machine power closely tracks duty cycle over time. However, interruptions can lead to periods of inactivity where machines are idle, creating gaps in job execution. To account for these idle periods, we employ a filtering strategy: only timestamps where all machines assigned to the run have a duty cycle of at least 80\% are included in the analysis. This filter ensures we capture machine power solely during active job execution, excluding periods of inactivity. For each run, the average on-duty machine power is calculated across all machines and all included timestamps.

This approach is applied to both complete and incomplete runs. Incomplete runs, which may be stopped for various reasons (e.g., sufficient steps processed for meaningful metrics), can still provide valuable data points, because the energy and Scope 2 carbon per step during completed steps are not affected by the eventual early termination of the workload.  We manually validate the calculated power for incomplete runs to ensure we only include runs that fall within a reasonable range.

\section{Propensity Score Weighting}
\label{appendix:Propensity}
Directly comparing emissions across TPU generations (e.g., CO$_{2}$e per machine-day) is challenging due to varying utilization patterns: older platforms generally exhibit lower utilization rates (both duty cycle and FLOPs/second) than newer platforms, potentially due to migration trends. Since each platform typically achieves higher performance-weighted energy efficiency at higher utilization, a naive comparison risks overstating the efficiency improvements of newer platforms. This liability comes from conflating two factors:
\begin{enumerate}
    \item Inherent efficiency gains: Newer platforms are inherently more energy efficient.
    \item Utilization differences: Newer platforms tend to operate at higher, more energy-efficient utilization rates.
\end{enumerate}
To isolate the impact of inherent efficiency gains (factor 1), we employ \textit{propensity score weighting}, a statistical technique ~\citep{Rosenbaum1983} that:
\begin{itemize}
    \item Balances utilization levels. By weighting observations based on their probability of being assigned to a specific accelerator given their utilization, we create a "pseudo-population" with similar utilization distributions across generations, and
    \item Reduces confounding. This balancing minimizes the confounding effect of utilization, allowing us to focus on the true impact of hardware improvements on \textcolor{black}{carbon intensity}.
\end{itemize}
We employ propensity score weighting to create a balanced comparison of machines at similar duty-cycle utilization levels. Specifically, we apply this method to two cohorts separately: Powerful TPUs (v4 and v5p) and Cost-efficient TPUs (v4i, v5e, and v6e (Trillium)) for comparisons within each cohort. Table \ref{tab:weighting-impact} shows how propensity score weighting balances duty cycle distributions across each class of TPUs. The propensity score weighting process involves:
\begin{enumerate}
    \item Calculating propensity scores. For each machine at a given time, determine the probability of the machine belonging to a specific TPU generation based on its duty cycle. Instead of looking at raw duty cycle rates, we'll group observations into levels.  Within each duty cycle level, we calculate what percentage of observations belong to each TPU generation, and that is the propensity score for this observation.
    \item Inverse probability weighting. For each observation, calculate the weight as the inverse of its propensity score for the accelerator it was actually assigned to. This gives higher weight to observations that were less likely to be assigned to their particular accelerator given their duty cycle rate. 
    \item Applying weights. Utilize the calculated propensity scores to weight the data points, effectively balancing the duty cycle distribution across generations. This assignment ensures that the comparison is not skewed by differences in duty cycle patterns. The formula is given as:
    \begin{equation*}
        \textrm{Weighted average for } M = \frac{\sum_i \textrm{weight}_i \times M_i}{\sum_i \textrm{weight}_i},
    \end{equation*}
    where $M_i$ is a metric of interest (e.g., average machine power and utilized FLOPs/second) for observation $i$, and $\textrm{weight}_i$ is the weight for observation i.
\end{enumerate}

\begin{table}[h]
    \caption{Results before and after applying propensity score weighting. 
    Red text highlights that duty cycles are equal after propensity score weighting.
    }
    \label{tab:weighting-impact}
    \centering
    \begin{tabular}{lccccc}
        \multicolumn{6}{c}{\textbf{Results before applying weighting}} \\
        \midrule
        Name & v4i & v5e & v6e & v4 & v5p \\
        Duty Cycle & 1 & 1.05 & 1.47 & 1 & 1.12 \\
        Utilized FLOPs/second & 1 & 1.16 & 7.49 & 1 & 2.47 \\ 
        Average Power & 1 &  0.99 & 1.96 & 1 & 1.90 \\ 
        Energy / ExaFLOP & 1 & 0.85 &  0.26 &  1 &  0.77 \\
        Carbon / ExaFLOP & 1 &  0.86 &   0.26 & 1 & 0.76 \\
        \midrule
        \multicolumn{6}{c}{\textbf{Results after applying weighting}} \\
        \midrule
        Name & v4i & v5e & v6e & v4 & v5p \\
        Duty Cycle & \textbf{\textcolor{black}{1}} & \textbf{\textcolor{black}{1}}& \textbf{\textcolor{black}{1}} & \textbf{\textcolor{black}{1}} & \textbf{\textcolor{black}{1}} \\
        Utilized FLOPs/second & 1 & 1.16 & 5.40 & 1 & 2.18 \\ 
        Average Power & 1 &  0.99 & 1.84 & 1 & 1.86 \\ 
        Energy / ExaFLOP & 1 & 0.85 &  0.34 &  1 &  0.85 \\
        Carbon / ExaFLOP & 1 &  0.86 &   0.34 & 1 & 0.84 \\
        \midrule    
        \end{tabular}
\end{table}

Here is an example to compute propensity scores and weights. Suppose only 30\% of the machines in one particular duty cycle level are v4, while 70\% are v5p.  The propensity scores for v4 and v5p machines would be 0.3 and 0.7. This score means v5p is over-represented in this duty cycle level. To account for this imbalance, we assign weights to each observation. Since v4 is less common, we give it a higher weight (3.3, inverse of its propensity score 0.3) compared to v5p (1.4, inverse of 0.7). This inversion compensates for v4 being underrepresented. These weights are then used when calculating the \textcolor{black}{carbon intensity}. By using weights, we ensure that the \textcolor{black}{carbon intensity} calculations are not skewed by the unequal representation of v4 and v5p in each duty cycle level in the data.

To demonstrate the benefits of propensity score weighting, the power, performance, and the resulting \textcolor{black}{carbon intensity} metrics, Table \ref{tab:weighting-impact} presents results before and after applying the methods. Relative values compared to v4i and v4 in each cohort are displayed in the table to facilitate comparisons among TPUs. Propensity score weighting makes the average duty cycles equal within each cohort. Using propensity score weighting, we create a more accurate (and more modest) estimate of the CCI improvement of newer generations by reducing confounding from duty cycle utilization. 


\bibliographystyle{ACM-Reference-Format}
\bibliography{references}

\end{document}